\documentclass[12pt,preprint]{aastex}

\shorttitle{Variable accretion and outflow in young brown dwarfs}
\shortauthors{Scholz \& Jayawardhana}

\begin{document}


\title{Variable accretion and outflow in young brown dwarfs}


\author{Alexander Scholz and Ray Jayawardhana}
\affil{Department of Astronomy \& Astrophysics, University of Toronto,
    60 St. George Street, Toronto, Ontario M5S3H8, Canada}
\email{aleks@astro.utoronto.ca}

\begin{abstract}
We report on the first dedicated monitoring campaign of spectroscopic
variability in young brown dwarfs. High-resolution optical spectra of
six targets in nearby star-forming regions were obtained over 11 nights
between 2005 January-March on the Magellan 6.5m telescope. We find
significant variability in H$\alpha$ and a number of other emission lines
related to accretion and outflow processes on a variety of timescales
ranging from hours to weeks to years. The most dramatic changes are seen for
2MASS J1207334-393254 (2M1207), 2MASS J11013205-7718249 (2M1101) and
ChaI-ISO217. We observe possible accretion rate changes by about an
order of magnitude in two of these objects, over timescales of weeks
(2M1207) or hours (2M1101). The accretion `burst' seen in 2M1101 could
be due to a `clumpy' flow. We also see indications for changes in the
outflow rate in at least three objects. In one case (ISO217), there
appears to be a $\sim$1-hour time lag between outflow and accretion
variations, consistent with a scenario in which the wind originates
from the inner disk edge. For some objects there is evidence for 
emission line variability induced by rotation. Our variability study 
supports a close to edge-on inclination for the brown dwarf LS-RCrA 1. 
The fact that all targets in our sample show variations in accretion and/or 
outflow indicators suggests that studies of young brown dwarf properties
should be based either on large samples or time series. As an example,
we demonstrate that the large scatter in the recently found accretion
rate vs. mass relationship can be explained primarily with variability.
The observed profile variations imply asymmetric accretion flows in
brown dwarfs, which, in turn, is evidence for magnetic funneling by
large-scale fields. We show that accreting sub-stellar objects may
harbor magnetic fields with $\sim$kG strength.
\end{abstract}


\section{Introduction}
\label{intro}

Brown dwarfs (BDs) are objects with masses intermediate between stars and planets. From
a variety of observational studies, there is now clear evidence that many 
BDs with ages $<10$\,Myr share properties with solar-mass classical T Tauri stars 
(CTTS), which are in a phase of active accretion. Young BDs have been found to 
harbour circum-sub-stellar disks, as indicated by near- and mid-infrared fluxes 
significantly higher than the photospheric fluxes from the object itself. Their 
spectral energy distribution in the infrared and in the sub-millimeter regime is nicely 
reproduced by disk models with either flat or flared disk geometry 
\cite[e.g.,][]{ntc02,jas03,pah03,mjn04}. Moreover, a large fraction of 
young BDs show the typical emission line spectrum of T Tauri stars, with broad, 
asymmetric H$\alpha$ lines, and additional emission lines like HeI and OI, 
which is direct evidence for ongoing accretion 
\cite[e.g.,][]{jmb02,jmb03,wb03,mhc03,ntm04,mjb05}. The accretion process is also 
believed to produce the large-amplitude, partly irregular photometric variability 
observed in young BDs \citep{se04,se05}. Finally, for some substellar objects 
outflows have been detected as forbidden line emission in [SII] 
and [OI] \citep{fc01,bmj04,bj04,wrb05}. Thus, it is quite clear that many 
young BDs undergo an accretion phase similar to more massive objects.

The general picture of accretion on T Tauri stars is the magnetospheric accretion model.
In this view, the accretion is funnelled by a large-scale magnetic field structure, which 
truncates the disk at a certain inner radius. The 
gas flows with nearly free-fall velocity from the disk to the star following the magnetic 
field lines and finally forms an accretion shock and hot spots near the atmosphere of the 
accreting object \cite[e.g.,][]{k91,sno94,hhc94}. Many recent studies provide ample support 
for this idea of magnetically channelled accretion, and there is growing evidence that
this scenario also applies to substellar objects.

Perhaps the most detailed view on the accretion process can be obtained by high-resolution 
optical spectroscopy. By analysing the emission line fluxes and profiles, it 
is possible to constrain the accretion parameters such as the 
infall rate or the temperature in the flow. Accreting objects, however, show a puzzling 
variety of emission line profiles, from nearly symmetric and double-peaked to classical 
or inverse P Cygni type features \citep[e.g.][]{foe95,rpl96}. Part of this diversity may be 
explained by different accretion flow geometries in 
individual objects. But we also have to take into account that we see the 
star-disk system in different projections. In the two extreme cases, the line of sight 
can point directly to the edge of the disk (edge-on view) or to the pole (face-on view). 
As a consequence, for different objects the received flux comes from different parts of 
the disk and the object, whereas other parts may be obscured.

On the other hand, it has long been known that T Tauri stars are variable on timescales
ranging from a few hours to years \citep[e.g.][]{hhg94,jb95,bca99,bga03}. Hot spots formed by 
the accretion flow co-rotate with the object, and might therefore not always be visible to 
the observer. Moreover, the rotational axis can be inclined with respect to the magnetic axis, 
resulting in a non-axisymmetric flow \citep{ruk03}. Simulations show that the magnetospheric accretion 
itself is unlikely to be stable over long timescales; reasons for instabilities include recurrent 
opening of the magnetic field lines \citep[e.g.][]{u02,mp05} and clumpy accretion \citep{mng93}. 
All these processes can lead to variability and, in particular, to variations in the emission 
line profiles. Summarising, rotation, accretion instabilities, geometry effects, and differences 
between individual objects are mainly responsible for the different types of emission line profiles 
observed for accreting objects.

From this reasoning it becomes clear that variability studies, in particular spectroscopic
monitoring, can be used to obtain a close-up view on the accretion behaviour of individual
objects, because they help to disentangle the different effects which influence the line
profiles. Indeed, from intensive monitoring campaigns for selected CTTS, it has been
possible to derive a detailed description of the accretion process for some objects 
\cite[e.g.][]{jb97,ab02,bga03}. At a minimum, variability studies provide 
information about the range of accretion parameters on a given object, allowing us an 
unbiased comparison between different objects. Thus, variability studies for individual 
targets are a valuable complement to the statistical analysis of large object samples. 

Therefore, we have carried out the first spectroscopic monitoring campaigns of young, 
accreting brown dwarfs. We selected targets which show broad and
asymmetric H$\alpha$ profile in previous studies, indicative of infalling
and/or outflowing material (see Sec. \ref{lines}). In addition, we aimed to cover a 
large variety of emission line profiles and thus accretion properties. Our final 
sample of six objects is given in Table \ref{targets}, which additionally contains 
aliases for all targets. In the remainder of the paper, we will use these short names.
For all six objects, the membership in nearby star-forming regions has been confirmed 
spectroscopically, and all have spectral types later than M6, indicating that they are 
either very close to or below the substellar limit \citep[see][]{lsj03}. For a more 
precise mass estimate for these objects, a better calibration of the stellar evolutionary 
tracks for young objects is required \citep[see][]{bca02}. However, since crossing the 
substellar limit is not expected to alter the accretion process in any way, we will 
conveniently call all targets 'brown dwarfs' in the rest of the paper, although not 
all of them may be bona fide substellar objects. 

First results from this study have already been published in \citet{sjb05} 
(hereafter SJB05), where we report on dramatic changes in the emission lines of the
BD 2M1207 (see Table \ref{targets}). In this paper, we give a consistent description of
the emission line variability for six targets, including 2M1207. In Sec. \ref{obs}
we present our observations and data reduction. Subsequently, we give a phenomenological 
description of the emission line spectra, including an analysis of the profile of the
most prominent feature, the H$\alpha$ line (Sec. \ref{Halpha}), and a survey of other 
features seen in our data (Sec. \ref{survey}). This information will then be used to 
constrain the accretion behaviour of our targets (Sect. \ref{disc}) and brown dwarfs in 
general (Sec. \ref{general}). Finally, we give our conclusions in Sec. \ref{conc}.

\section{Observations and data analysis}
\label{obs}

\subsection{Observation and basic reduction}

All spectra for this study have been taken with the MIKE spectrograph at the 
Magellan-Clay 6.5\,m telescope on Las Campanas. MIKE is a double echelle instrument, 
consisting of blue and red arms, providing coverage from 3400 to 5000\,\AA~ and 4900 
to 9300\,\AA, respectively, in our configuration. We obtained most of the time series 
spectra for our six targets in an observing run from 17 to 19th of March 2005. Except 
for ChaH$\alpha$1, at least one additional spectrum per target was taken in a second 
observing run at the end of March. Three extra spectra for 2M1207 were obtained in a 
preceding campaign in January-February 2005. In total, the variability study is based on 
13 spectra for 2M1207, seven for 2M1101, five for $\rho$Oph-32, ISO217, ChaH$\alpha$-1, 
and four for LS-RCrA\,1 (see Table \ref{log}). In the main observing run, we used on-chip 
binning of $2\times 2$ pixels in combination with slit sizes of 1\farcs0 or 0\farcs7, 
resulting in a resolution of $R \sim 20000$. All exposures were split into three single 
spectra to avoid excessive contamination by cosmic rays. The two auxiliary 
runs had the primary goal to measure high-precision radial velocities, therefore we aimed 
for the highest possible resolution. For that reason, spectra from these runs were taken 
with a 0\farcs35 slit in combination with either no or $2\times 2$ binning. Therefore, 
the data from our primary run have in general a higher signal-to-noise ratio (hereafter s/n), 
but lower resolution. The exposure times for our targets were between 15 and 60\,min. In 
Table \ref{log} we list for all spectra the observing times, configurations, and exposure 
times. In all runs, we additionally observed a spectrophotometric standard star for order 
definition and (relative) flux calibration.

The basic reduction of all spectra was done using standard routines in the {\it doecslit}
package within IRAF\footnote{IRAF is distributed by National Optical Astronomy Observatories, 
which is operated by the Association of Universities for Research in Astronomy, Inc., under 
contract with the National Science Foundation.}. The procedure includes overscan subtraction,
flatfield correction, sky background subtraction, order extraction, and blaze correction
using the standard star spectrum. The wavelength calibration was carried out using
Thorium-Argon spectra; we obtained a typical precision of a few m\AA~for the wavelength 
solution. Cosmic rays were removed by applying a small-scale median filter to each order, 
comparing the filtered spectrum with the original data, and replacing all outliers with the 
median value. In cases where we split the total exposure time to several spectra, these single 
exposures were co-added after the extraction. 

\subsection{Line measurements}
\label{lines}

Our spectra show a variety of emission lines, which are believed to form either
by accretion, outflows/winds, or magnetic activity. We selected a set of features,
which should allow us to disentangle the effects of these three processes, and
measured equivalent widths (hereafter EW) for all of them.

The dominant emission line in all spectra is the H$\alpha$ feature at 6562.8\AA, which has 
its origin in the chromosphere and/or in the accretion flow. To decide whether an
object is accreting or not, we take advantage of the fact that we expect 
the material in the accretion flow to be moving close to free-fall velocity, which is 
$\sim 150\,\mathrm{km s^{-1}}$, whereas the velocity for chromospheric emission 
is usually significantly smaller. Based on these considerations, \citet{wb03} and
\citet{jmb03} introduced the 10\% width of the full peak height (hereafter 10\% width)
as a useful measurement to discriminate accretors and non-accretors. \citet{jmb03}
suggested a threshold of $\sim 200\,\mathrm{km s^{-1}}$ for substellar objects, and
\cite{mjb05} demonstrate that this is a reasonable value to separate accreting
and merely chromospherically active objects. It turned out that the 10\% widths are
indeed nicely correlated with the accretion rates derived from conventional
veiling analysis \citep{ntm04}. It can therefore be used to obtain an accretion
rate estimate from the emission line profile, which is particularly useful for BDs,
where the continuum is faint. It should be noted, however, that an extremely active 
and fast rotating object might be able to produce a 10\% width of about
$200\,\mathrm{km s^{-1}}$ or more. From our targets, however, none is known to
be an exceptional fast rotator. The $v \sin i$ values derived so far (see Table
\ref{targets}) indicate rotational velocities $<20\,\mathrm{kms^{-1}}$.

The profile of H$\alpha$ can be used as an additional indicator for ongoing accretion: 
Chromospheric H$\alpha$ profiles are in most cases more or less symmetric, and often 
a small central dip due to self-absorption in the chromosphere is seen. On the other hand, 
accreting objects usually exhibit strongly asymmetric profiles, with broad wings and either 
red- or blue-shifted absorption features, indicative of high-velocity infalling or 
outflowing material. In the following, we will therefore use the 10\% width and the 
shape of the profile to distinguish between non-accretors and accretors.

All our targets show emission in the higher Balmer lines H$\beta$ and H$\gamma$,
which require higher temperature regions in comparison with H$\alpha$. Again, the
width of the lines can be used to discriminate accretors and non-accretors. Additionally,
many spectra show HeI, most prominently at 5875 and 6678\,\AA, OI at 8446\,\AA,
and the CaII triplet in emission. If detected, we measured EW for these lines. The 
HeI5875 feature is known to be in absorption in many chromospherically active stars, 
but it can also appear in emission during flares in active M dwarfs 
\citep[and references therein]{sho97}. Although young BDs are known to show very 
few flares \citep[see][]{se05}, we cannot exclude that the HeI5875 emission has its 
origin in the chromosphere. HeI6678, OI8446, and the CaII triplet, however, appear 
preferentially in accretors, as argued by \citet{mjb05}. Therefore we use these
lines as complementary information to distinguish between accretors and non-accretors.
Particularly the HeI6678 EW can be used as a strong discriminator between
accretors and non-accretors: \citet{grh02} observed 676 M dwarfs, most of them active, 
in the solar neighbourhood, and although many of them have HeI6678 in emission, none has 
an EW larger than 0.25\,\AA. For our spectra, this is lower than the detection limit, 
i.e. if we detect HeI6678 in emission, it is very likely caused by accretion.

Similarly to H$\alpha$, many emission lines of accreting objects can be formed in
infalling as well as in the chromosphere or in outflowing material, as has been shown 
at least for HeI \citep{bek01} and CaII \citep[see][]{fc01}. The line ratio in the CaII triplet,
however, gives a first indication of whether the flux comes mainly from infall or 
outflow: For low-density outflowing material, we expect line ratios of 1:9:5, as it
has been observed for Herbig Haro objects \citep{rbg86}. On the other hand, line
ratios close to 1:1:1 imply saturation and optically thick CaII emission, as argued by 
\citet{gh88}, and is therefore evidence for an accretion-related origin.

Some objects show clear evidence for forbidden line emission, particularly in
[SII] (6730\,\AA) and [OI] (6300\,\AA). If possible, we measured EW for these lines
as well. Forbidden line emission is a clear indication for mass outflow, since these
lines can only form in low-density regions, and not in the accretion shock \citep[see][]{hms97}. 
They can therefore also be used to decide whether the permitted emission lines are mainly
formed by outflow or infall.

For all line measurements, the spectra were shifted to the local stellar rest frame.
We used literature radial velocities for 2M1207, ChaH$\alpha$1, and LS-RCrA 1 (see 
Table \ref{targets}). The uncertainties of these measurements are at most $3\,\mathrm{km s^{-1}}$.
For the remaining three targets, there is no previous radial velocity measurement. Using the 
Li6708 absorption feature, we are able to derive an estimate of $0\pm 3\,\mathrm{km s^{-1}}$
for $\rho$Oph-32, which is roughly consistent with the velocity of the cloud and of young stars 
in $\rho$\,Oph \citep[see][]{dgc05}. 2M1101 and ISO217 belong to the ChaI star forming region, 
where the brown dwarfs have been found to have average radial velocities of 
$\sim 15\,\mathrm{km s^{-1}}$ with only a small dispersion of 2-3$\,\mathrm{km s^{-1}}$ 
\citep{jg01,j05}. For these two targets, we therefore adopt the ChaI average value as radial 
velocity. Since the radial velocity error is the main source of uncertainty in velocity, 
we estimate our error for all velocity values to be $\sim 3\,\mathrm{km s^{-1}}$, still 
lower than our spectral resolution.

\subsection{Continuum analysis}
\label{cont}

Our general observing strategy was to keep the exposure time as short as possible, to be able
to obtain more than one spectrum per night for each of our target. As a consequence, the
s/n in the continuum is low, particularly for $\lambda<7000$\,\AA, deterring us from a 
detailed analysis of the photospheric lines for our objects. Many important emission 
lines are in regions with faint continuum, e.g. the 
complete Hydrogen Balmer series, and we have to make sure that the continuum emission
is dominated by the flux from the target, and not significantly contaminated by imperfect 
background or flatfield correction. We compared the averaged spectral energy distribution 
(SED) in our data with model spectra from \citet{ahs00}, selected to match effective 
temperature and gravity of our targets. Both the observed and the theoretical SED were 
normalised to the flux at 8600\,\AA, where we safely detect the continuum in all our spectra with 
s/n $>10$, and the ratio between observed and modeled spectrum was computed. This ratio is expected 
to scatter around unity as long as the observed continuum is determined predominantly by the 
photospheric flux from the target. For all our targets we find good agreement between observed 
and model SED for $\lambda>4800$\,\AA. As expected, the limit for a reliable continuum detection 
is a function of s/n. We conclude that all line measurements down to the H$\beta$ line
at 4861\,\AA~are reliable.

Since all our targets are known to accrete, they might show excess continuum emission
due to the so-called veiling, which is produced in the accretion funnel and 'fills' the
photospheric absorption lines. Strong veiling is often seen in accreting T Tauri stars, and
is known to correlate with the accretion rates \citep[e.g.,][]{bb90}. Since the veiling
is superimposed on the photospheric flux, it may prevent a reliable continuum measurement and
thus influence the equivalent widths. Recent studies, however, have shown that the accretion 
rates steadily decrease with the mass of the central object. As a consequence, the average 
accretion rates of brown dwarfs are in the range of $10^{-11}\,M_{\odot}\mathrm{yr}^{-1}$ 
\citep{mhc03,ntm04,mjb05}. Even in our sample of strongly accreting brown dwarfs, the mean 
accretion rate is probably only $10^{-10}\,M_{\odot}\mathrm{yr}^{-1}$. At these low accretion 
levels the veiling is safely below 10\% at least for $\lambda>5500$\,\AA~\citep{wb03,mhc03}, and 
thus negligible (and not detectable) in the red part of our spectra.

Veiling is a strong function of wavelength and in most cases becomes stronger in the blue spectral 
range. At 5000\,\AA~it can be twice as strong as at 7000\,\AA, as shown for CTTS \citep{ab02,bb90}. 
Discrepancies between photospheric and observed SED for $\lambda<4800$\,\AA, as we have detected 
in some cases, might therefore be caused by veiling, although this is unlikely given the fact that 
these discrepancies are strongest in spectra with very poor s/n, which clearly points to an 
instrumental effect. Therefore, our line measurements down to H$\beta$ are probably not significantly 
affected by veiling. 

\section{H$\alpha$ variability}
\label{Halpha}

\begin{figure*}[t]
\begin{center}
\includegraphics[width=4.0cm]{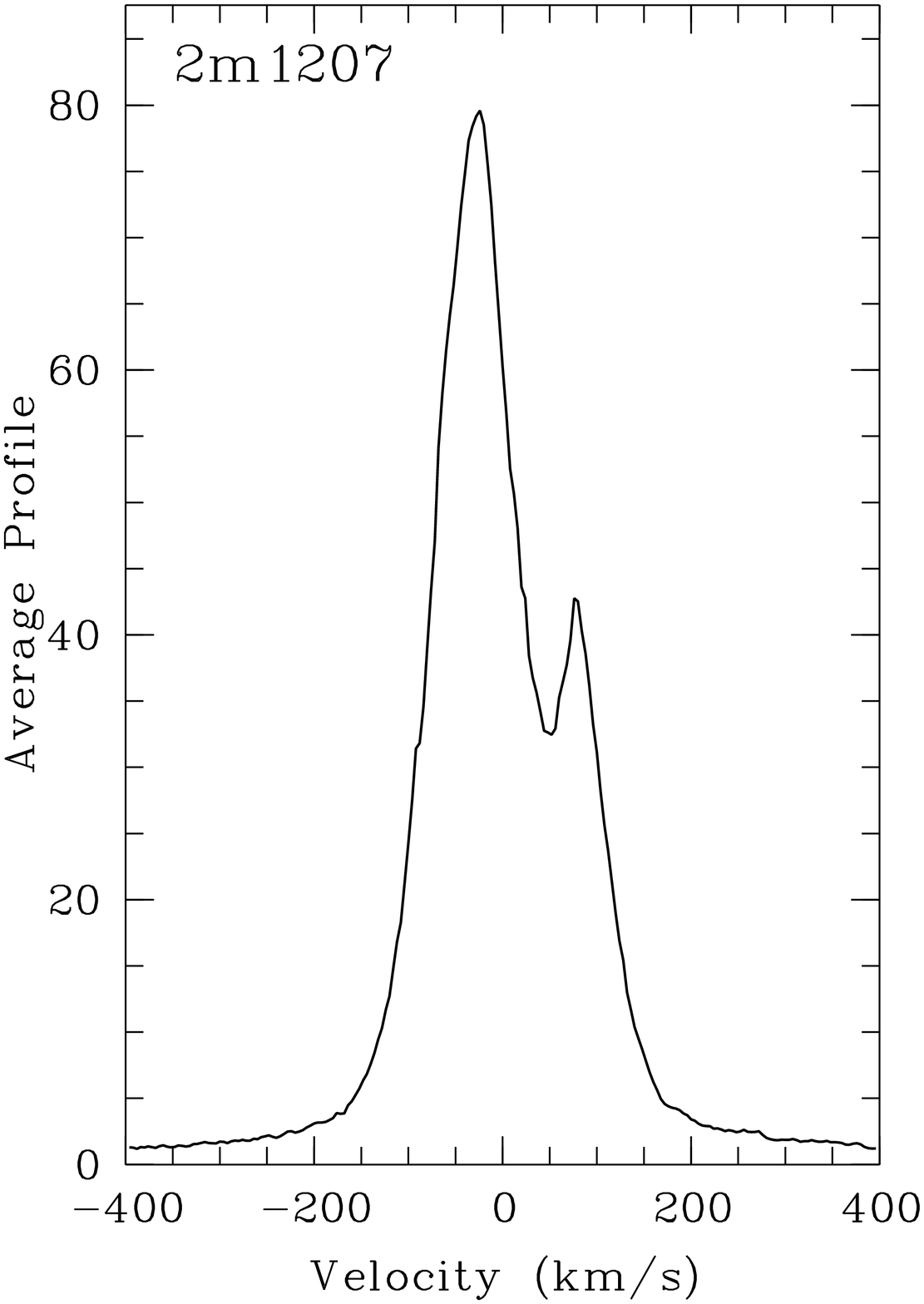}
\includegraphics[width=4.0cm]{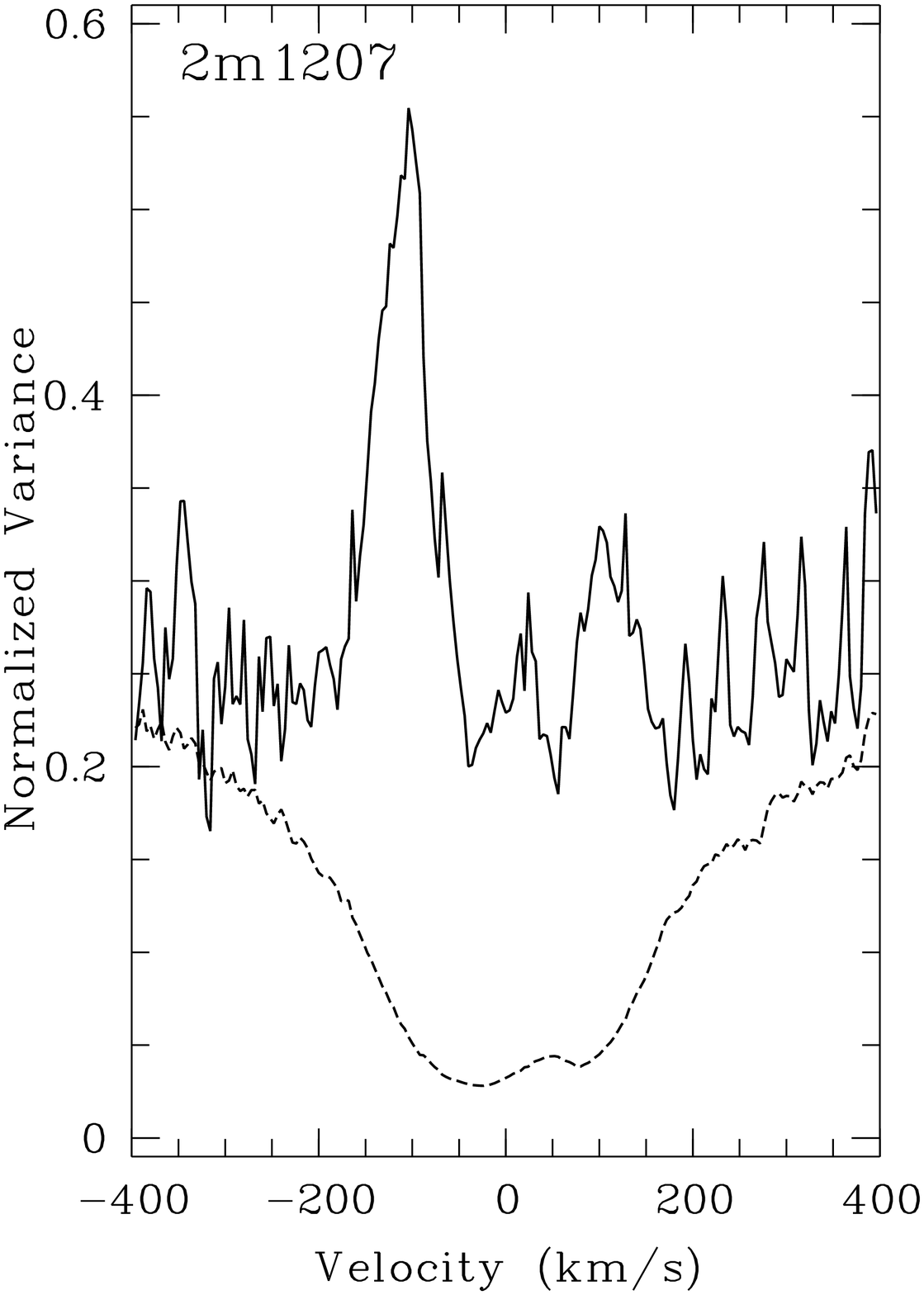}
\includegraphics[width=4.0cm]{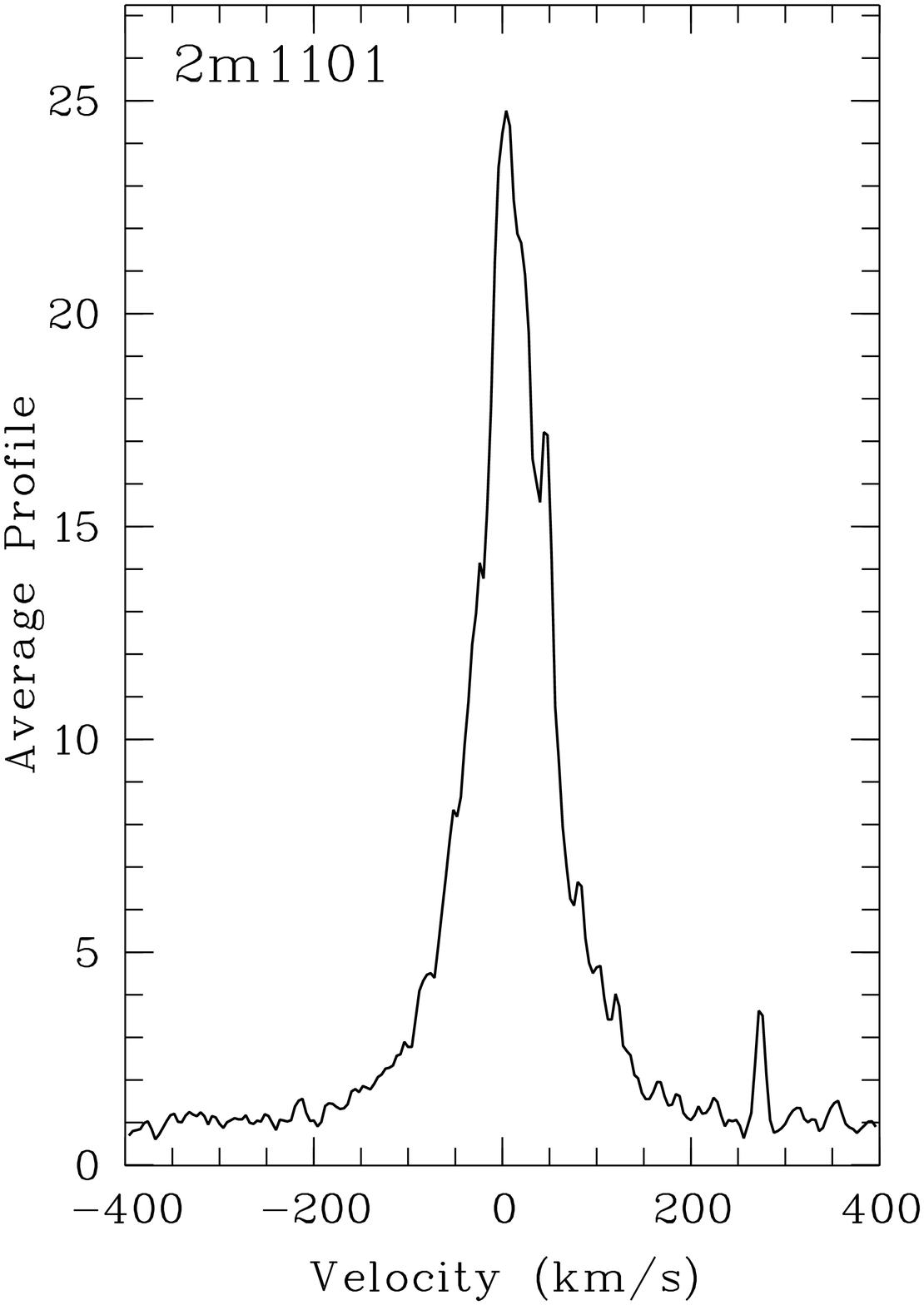}
\includegraphics[width=4.0cm]{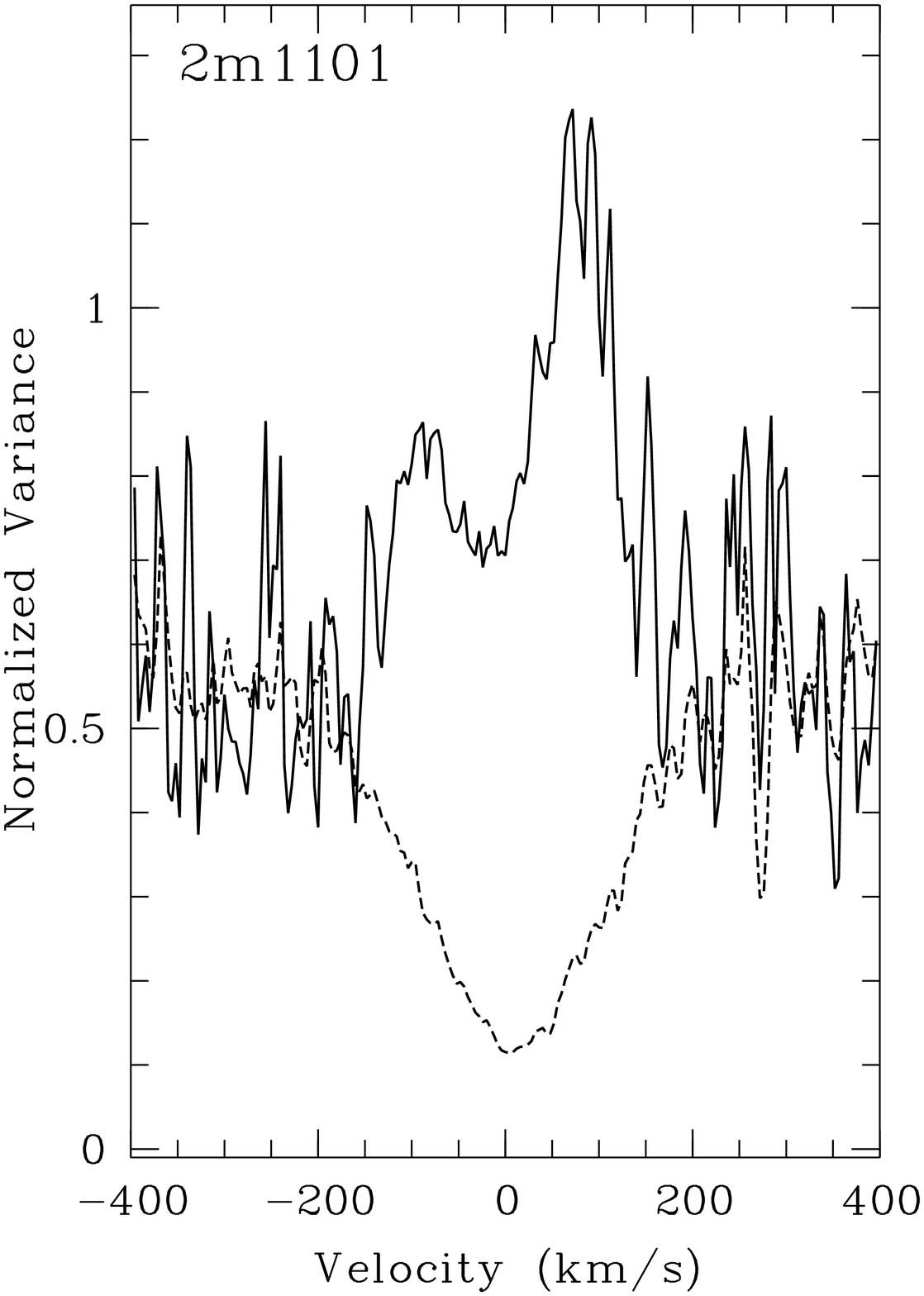}\\
\includegraphics[width=4.0cm]{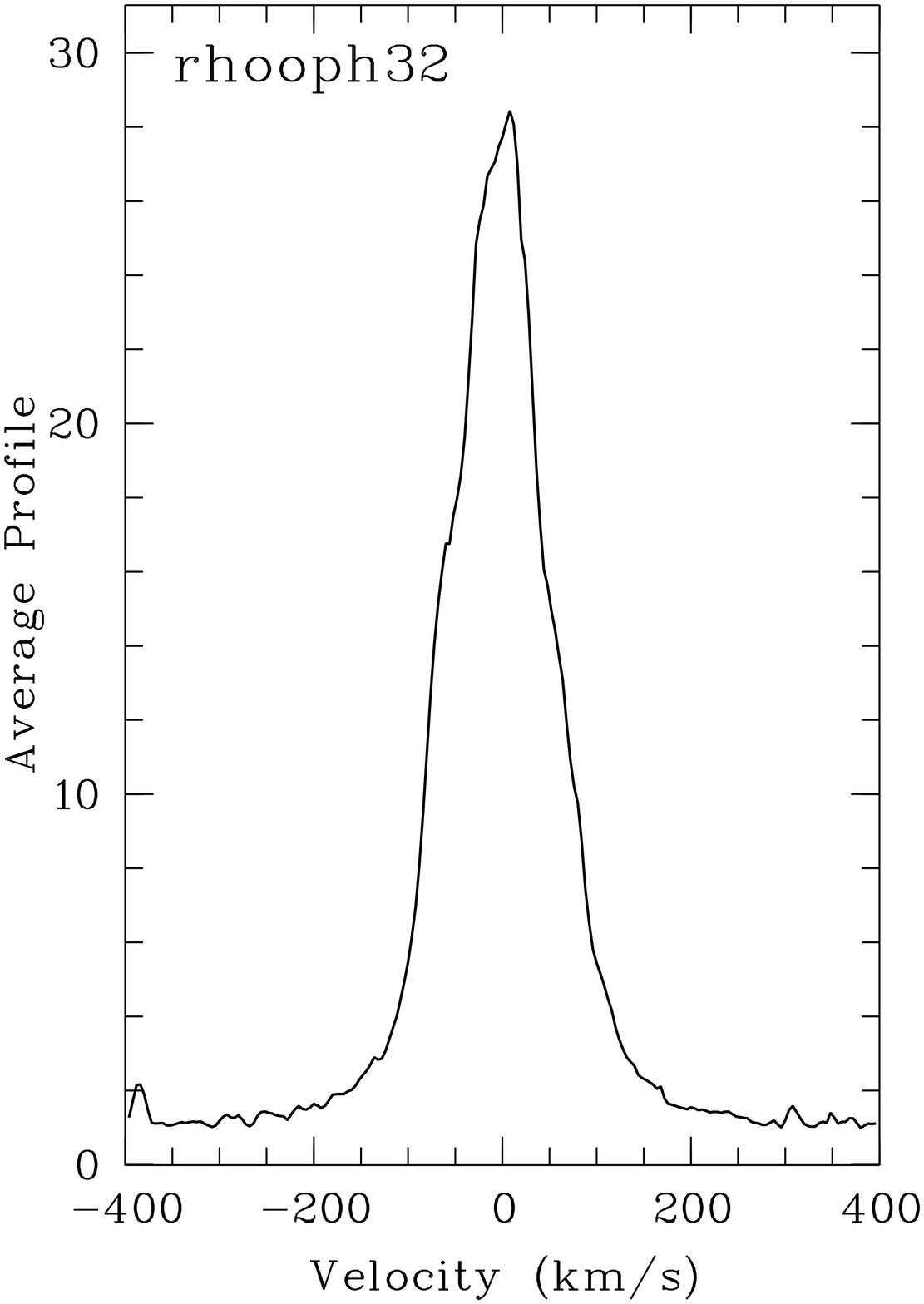}
\includegraphics[width=4.0cm]{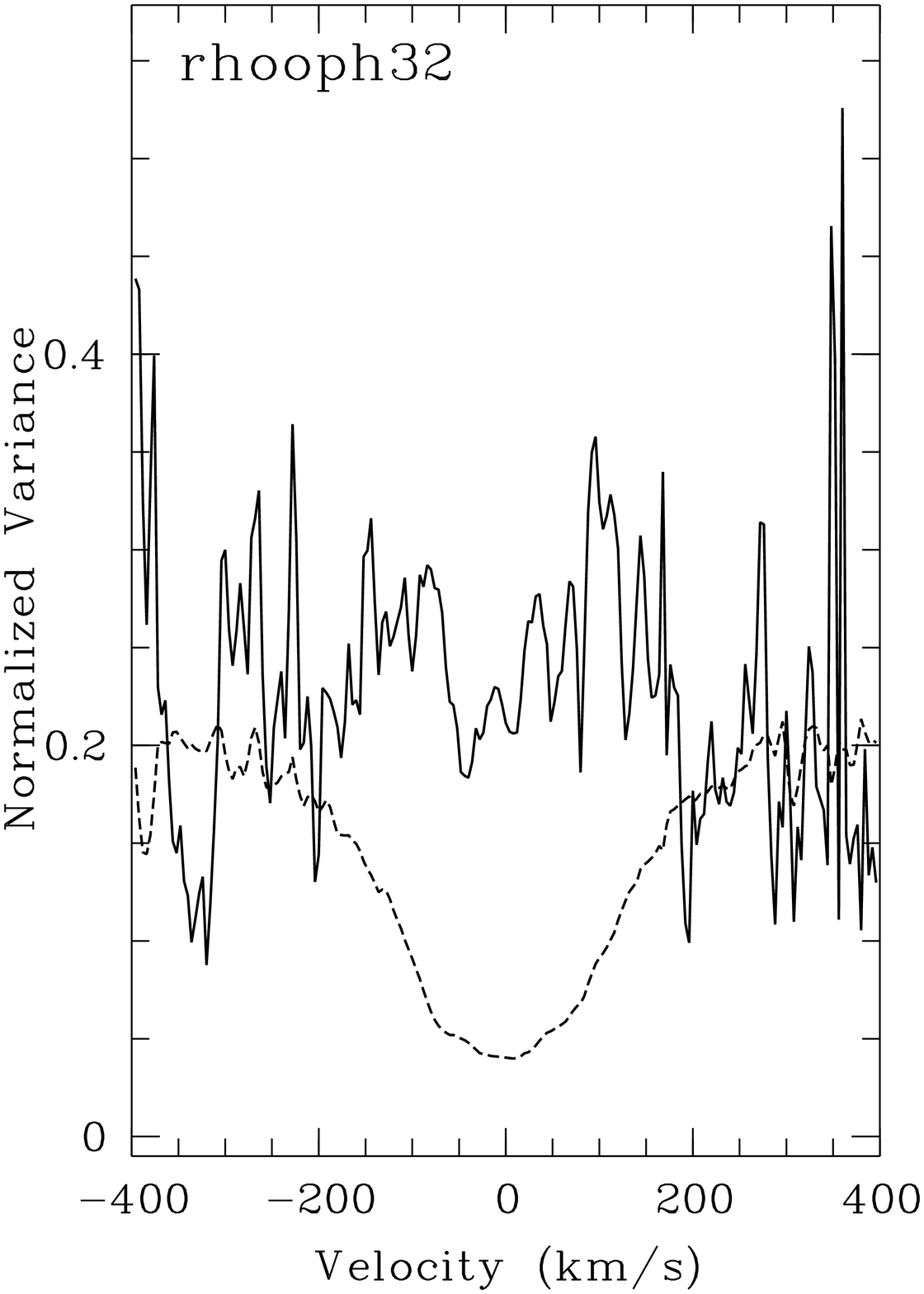}
\includegraphics[width=4.0cm]{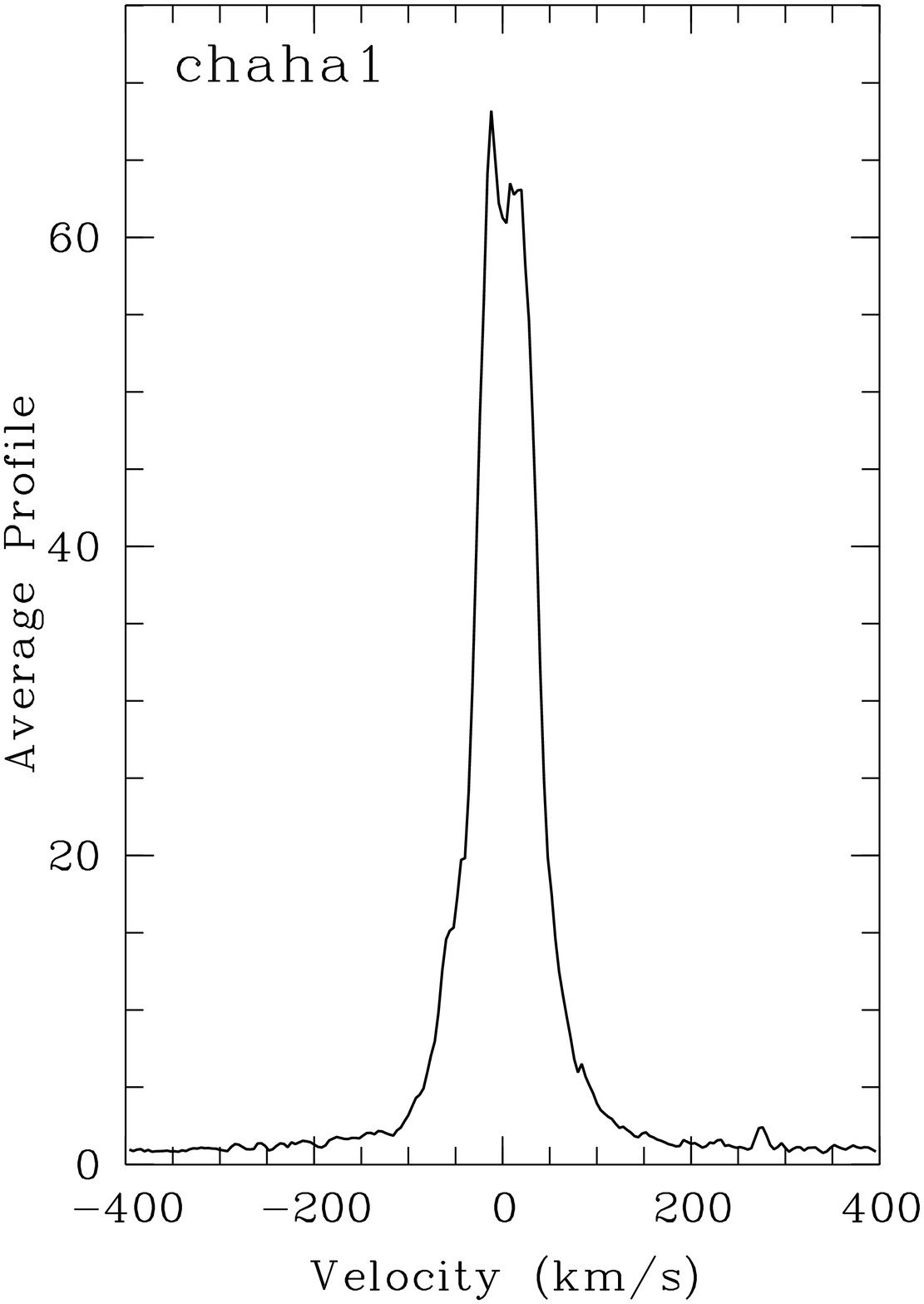}
\includegraphics[width=4.0cm]{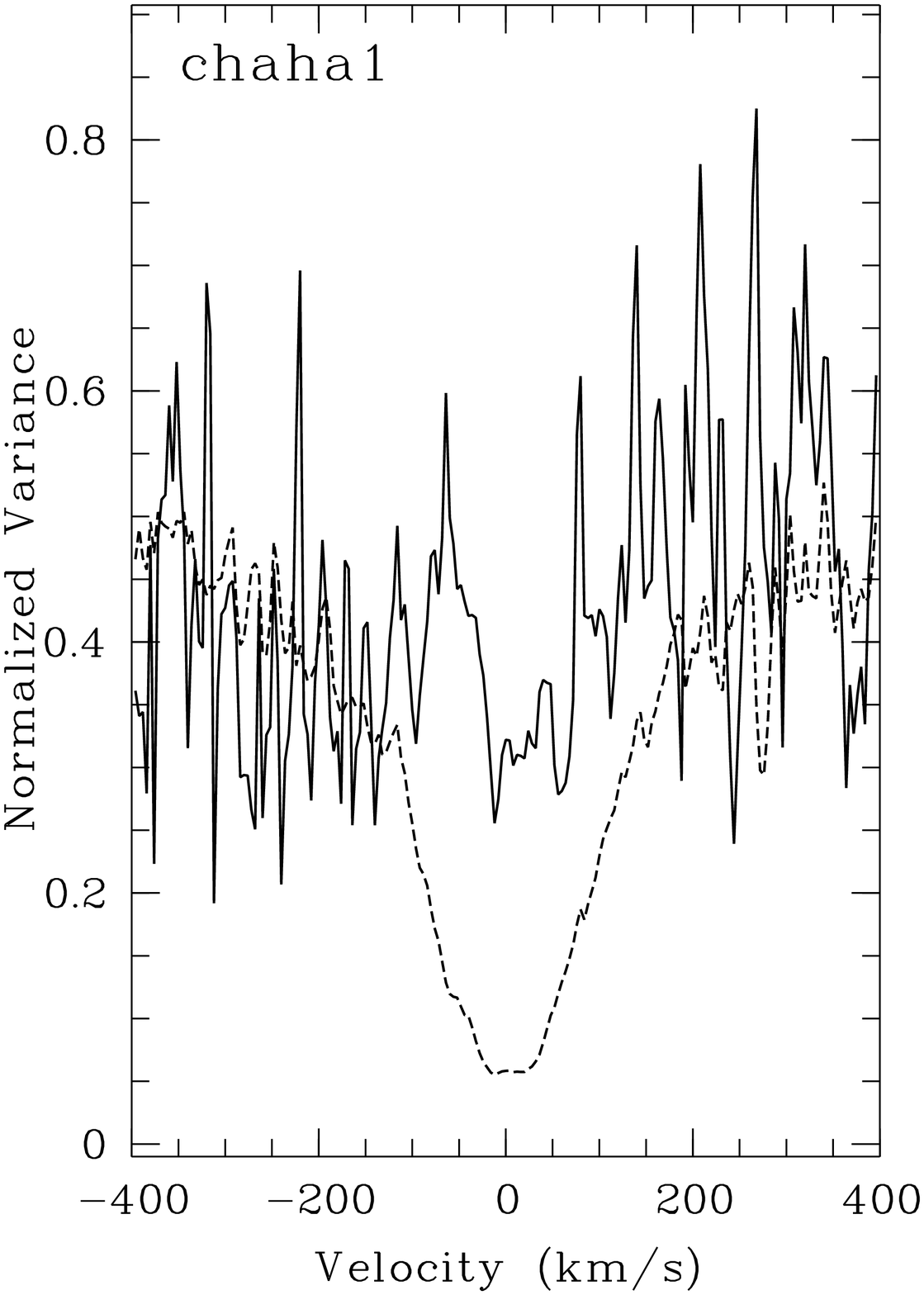}\\
\includegraphics[width=4.0cm]{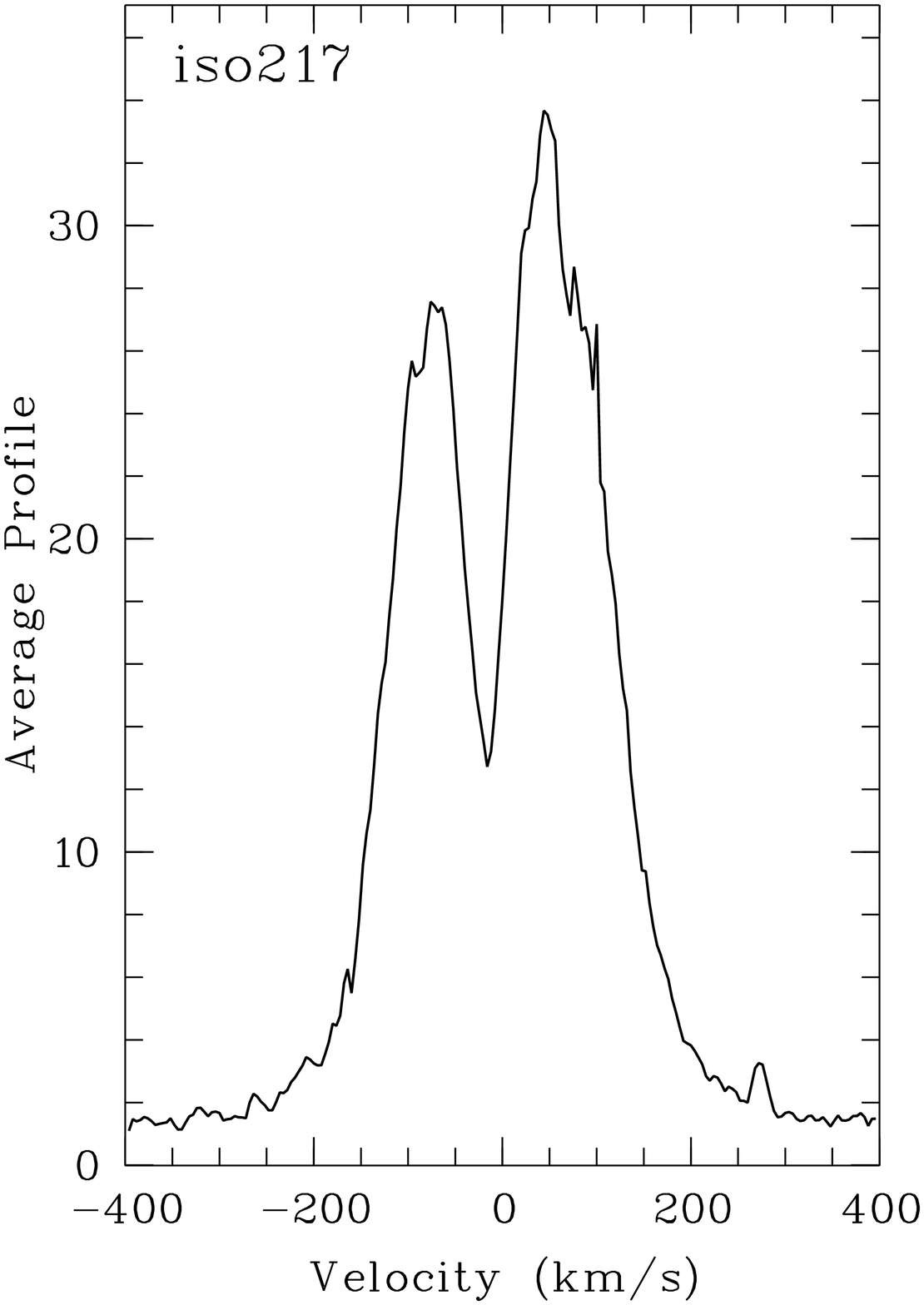}
\includegraphics[width=4.0cm]{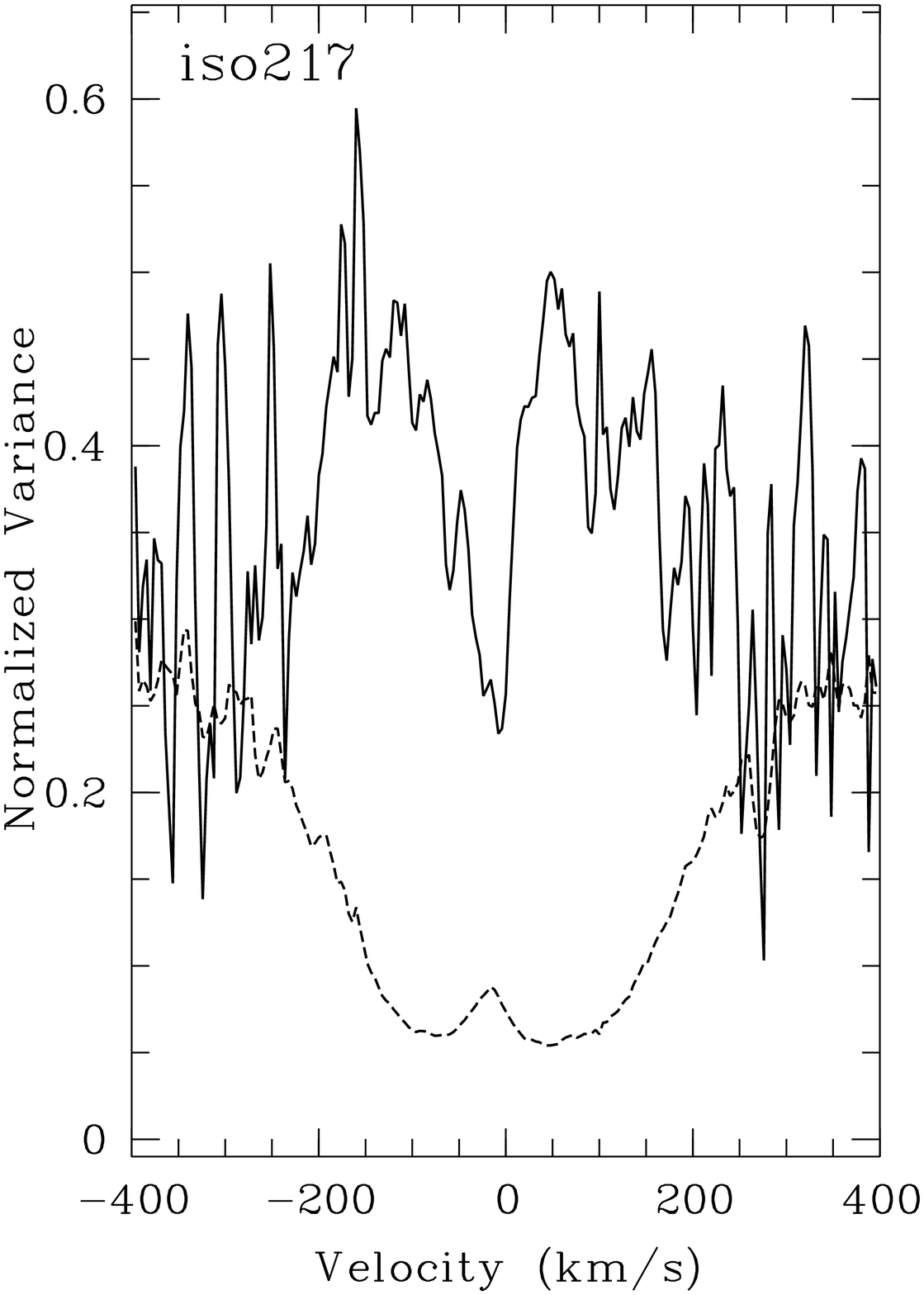}
\includegraphics[width=4.0cm]{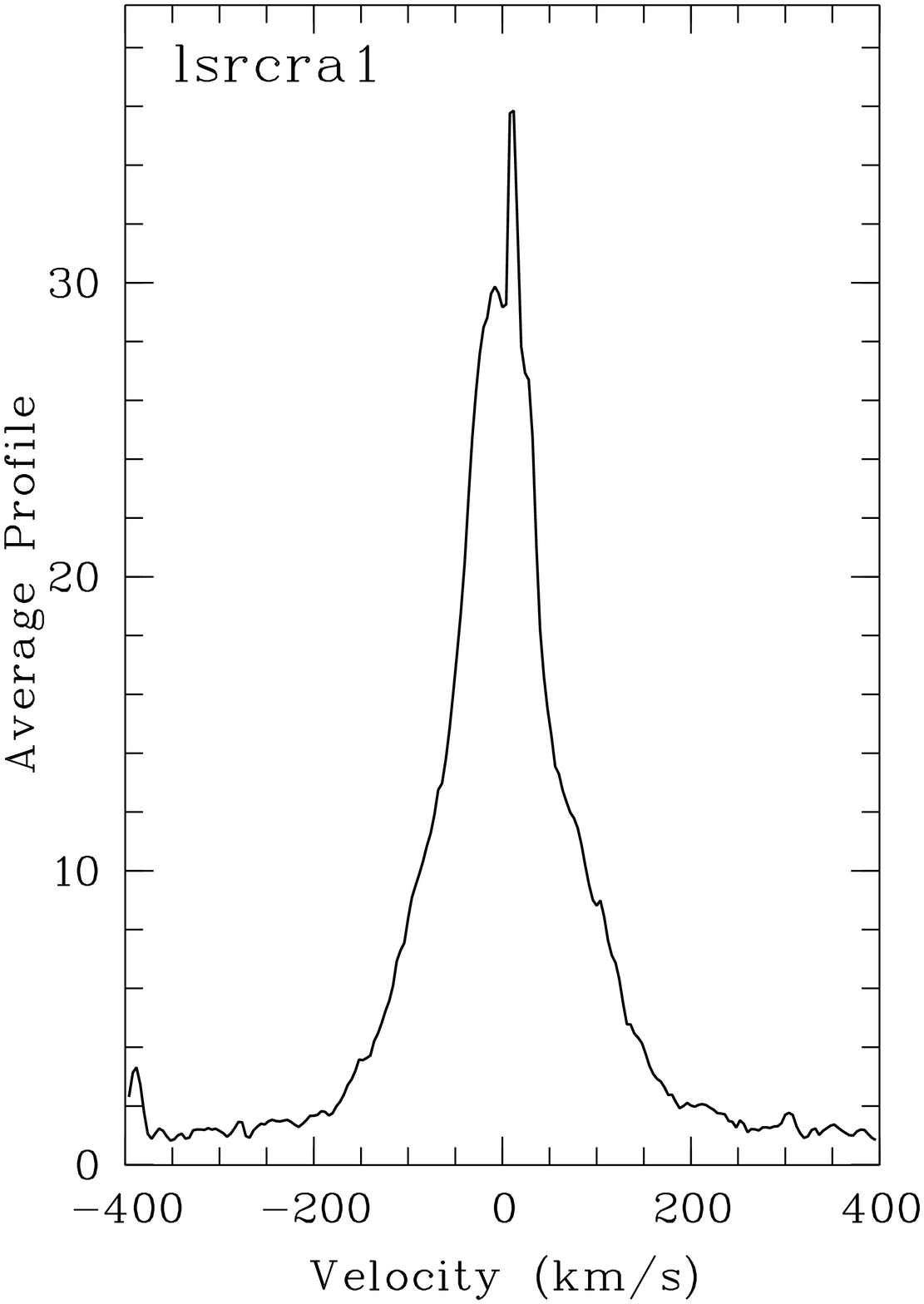}
\includegraphics[width=4.0cm]{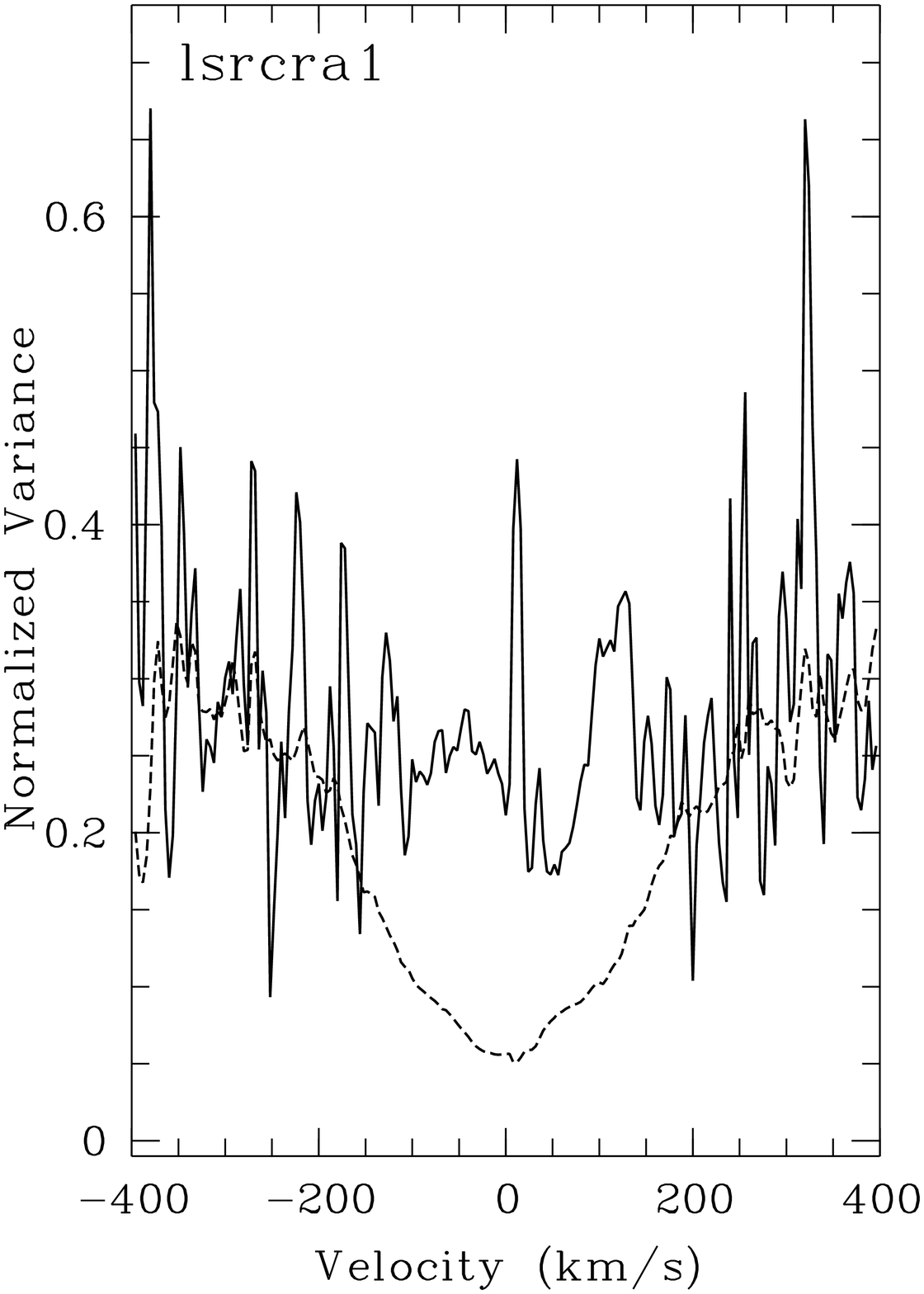}\\
\caption{Average and normalised variance H$\alpha$ profiles. Dashed lines show the level of zero
variability (see Sect. \ref{Halpha} for explanation).\label{avvar}}
\end{center}
\end{figure*}

The dominant emission feature for all targets is the H$\alpha$ line at 6562.8\AA. Our 
high-resolution spectra allow us a detailed analysis of the line profile. In Fig. 
\ref{avvar} we present the average H$\alpha$ profile $\overline{I}(v)$ for all six targets,
calculated over the complete time series. All single spectra were normalised to the continuum
level before the averaging. In addition, we derived the normalised
variance profile $\sigma_N^2(v)$ following \citet{jb95} by calculating the variance in in 
each velocity bin and normalising by $\overline{I}(v)$. This allows us to quantify the 
variability as a function of velocity. Both profiles are determined by averaging
over a moving velocity window with a width of $10\,\mathrm{km s^{-1}}$; the stepsize 
is $1\,\mathrm{km s^{-1}}$. 

To characterise the total amount of variability, \citet{jb95} used the EW of the 
variance profile, i.e. the integral over $\sigma_N^2(v)$. If we use this criterion,
three of our targets are highly variable (2M1207, 2M1101, ISO217), but for the remaining
three the H$\alpha$ profile is more or less constant. Using the EW variance as
variability measurement implicitly assumes that the zero variability level is constant
with velocity, which is usually the case for stars with strong continuum. In our spectra,
however, the s/n changes drastically with velocity; it is much higher in the core of H$\alpha$ 
than in the wings and the continuum. As can be seen from Fig. \ref{avvar}, the normalised
variance in the continuum ($\sigma_{N,0}^2$) is in the range of 0.2-0.5 for our targets,
implying a s/n of 1.5-2.5 in the continuum. Assuming Poisson noise, the expected 
$\sigma_N^2(v)$ without any variability (i.e. the zero variability level) can be 
estimated as $\sigma_N^2(v) = \sigma_{N,0}^2 \times (\sqrt{\overline{I}} / \overline{I})^2$. 
This function is shown with dashed lines in Fig. \ref{avvar}. 
As expected, the continuum is non-variable (by definition), because all spectra were 
normalised to the continuum before averaging. Using this zero-variability level, it is 
now clear that all six targets show significant variability in the profile. As a better 
way to assess the total variability in H$\alpha$ we define the H$\alpha$ variability index 
(H$\alpha$VI), calculated by integrating the difference between variance profile and zero 
variability level between -400 and $400\,\mathrm{km s^{-1}}$. These values are given in 
the last column of Table \ref{targets}. 

Our targets show a large variety of H$\alpha$ profiles, as apparent from the average
profiles. The variance profiles exhibit in most cases a blue-shifted and a red-shifted 
peak at velocities between 50 and $150\,\mathrm{km s^{-1}}$, sometimes a third central 
peak is visible. Three of our objects (2M1207, 2M1101, ISO217) show strong variability 
in this line with H$\alpha\mathrm{VI}>25$, whereas the remaining three targets are less 
variable. The variance profiles are in most cases quite different from the 
average profiles, demonstrating that the line variability is not only due to changes in 
the underlying continuum, because in this case the two profiles would be very similar
\citep[see][]{jb95,ab02}. 

2M1207 shows clear indications for an asymmetry on the red side, most likely caused by 
red-shifted absorption features in parts of the time series (see SJB05). The variance 
profile clearly has three peaks, where most of the variability is produced in the blue 
wing of the line. 2M1101 has only a minor asymmetry in the profile with a hint of a 
red-shifted second peak. This is surprising, given the fact that the only available 
literature spectrum for this object shows a P Cygni type H$\alpha$ profile with very strong 
absorption in the blue wing \citep{mlb05}. In contrast to 2M1207, here the variability 
in H$\alpha$ is mainly in the red part of the line. ISO217 has a double-peaked profile 
in all spectra, where the blue peak is significantly weaker than the red one. Both peaks also 
appear in the variance profile, but here they have about the same intensity. The 
remaining three targets have a single-peak profile, and the variance profiles do not show 
strong features. ChaH$\alpha$\,1 and LS-RCrA 1 have a small central absorption dip, whereas 
$\rho$Oph-32 has a slight asymmetry on the blue side, which may be due to a blue-shifted 
absorption feature.

\begin{figure*}
\begin{center}
\includegraphics[width=4.0cm]{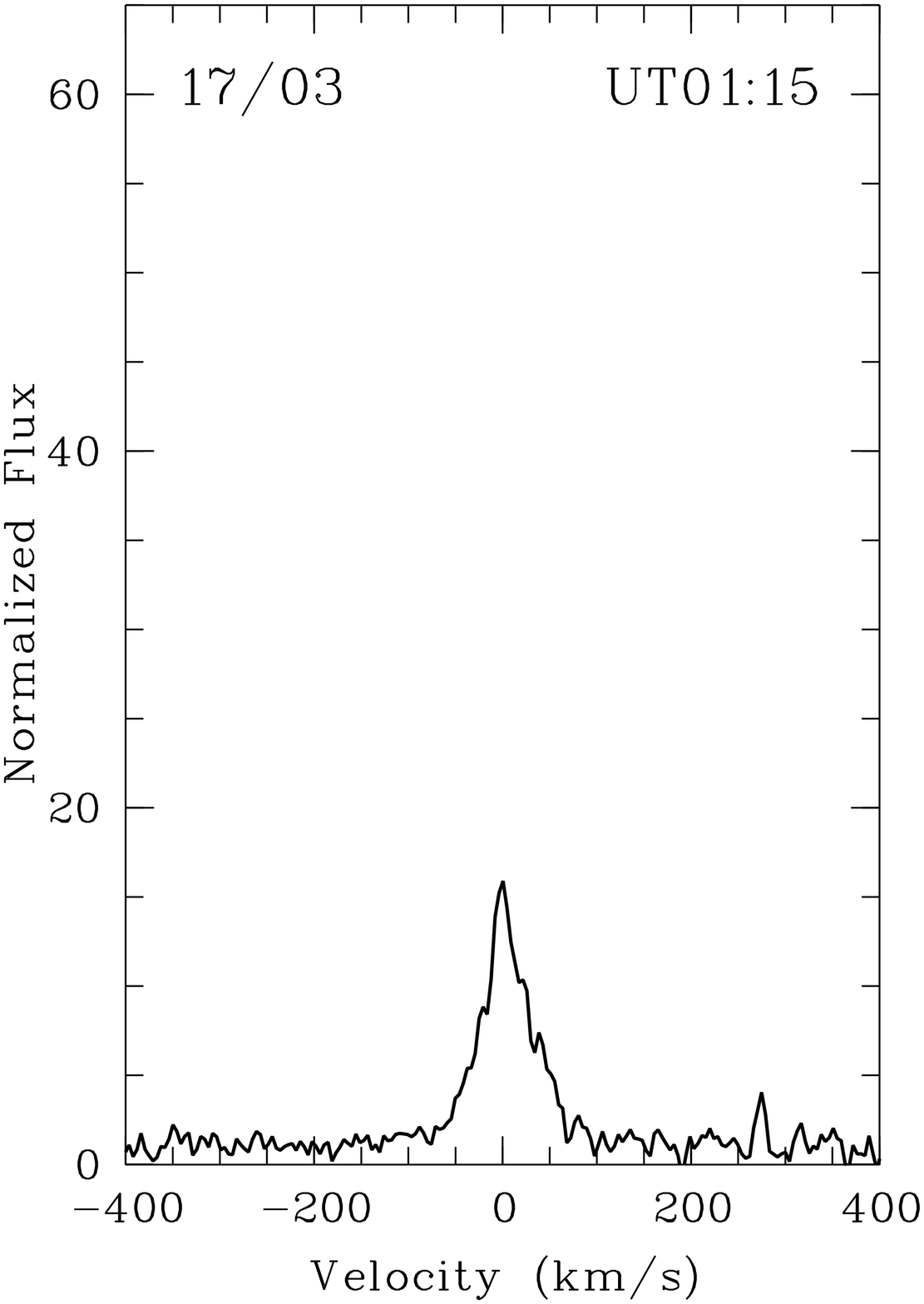}
\includegraphics[width=4.0cm]{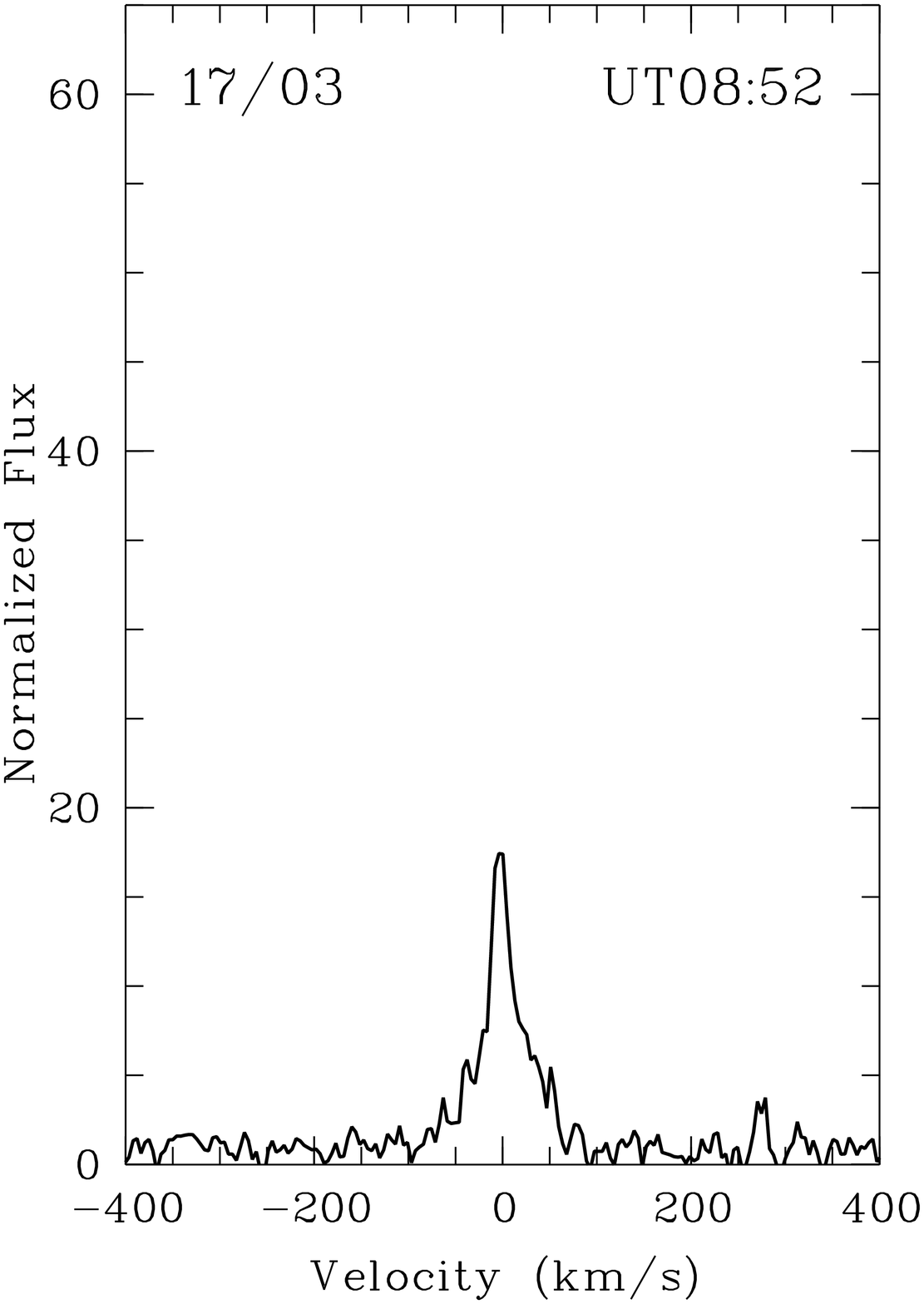}
\includegraphics[width=4.0cm]{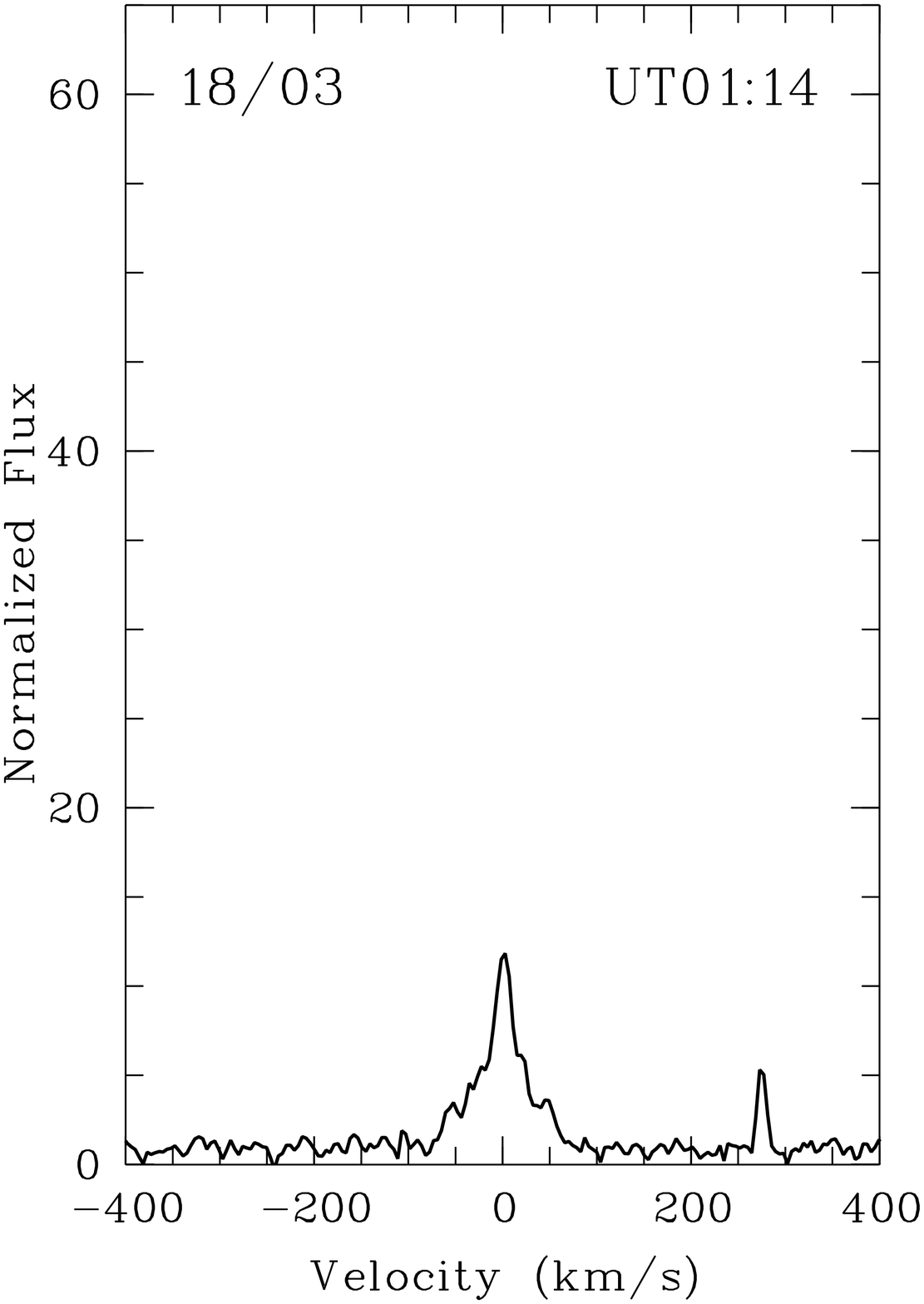}
\includegraphics[width=4.0cm]{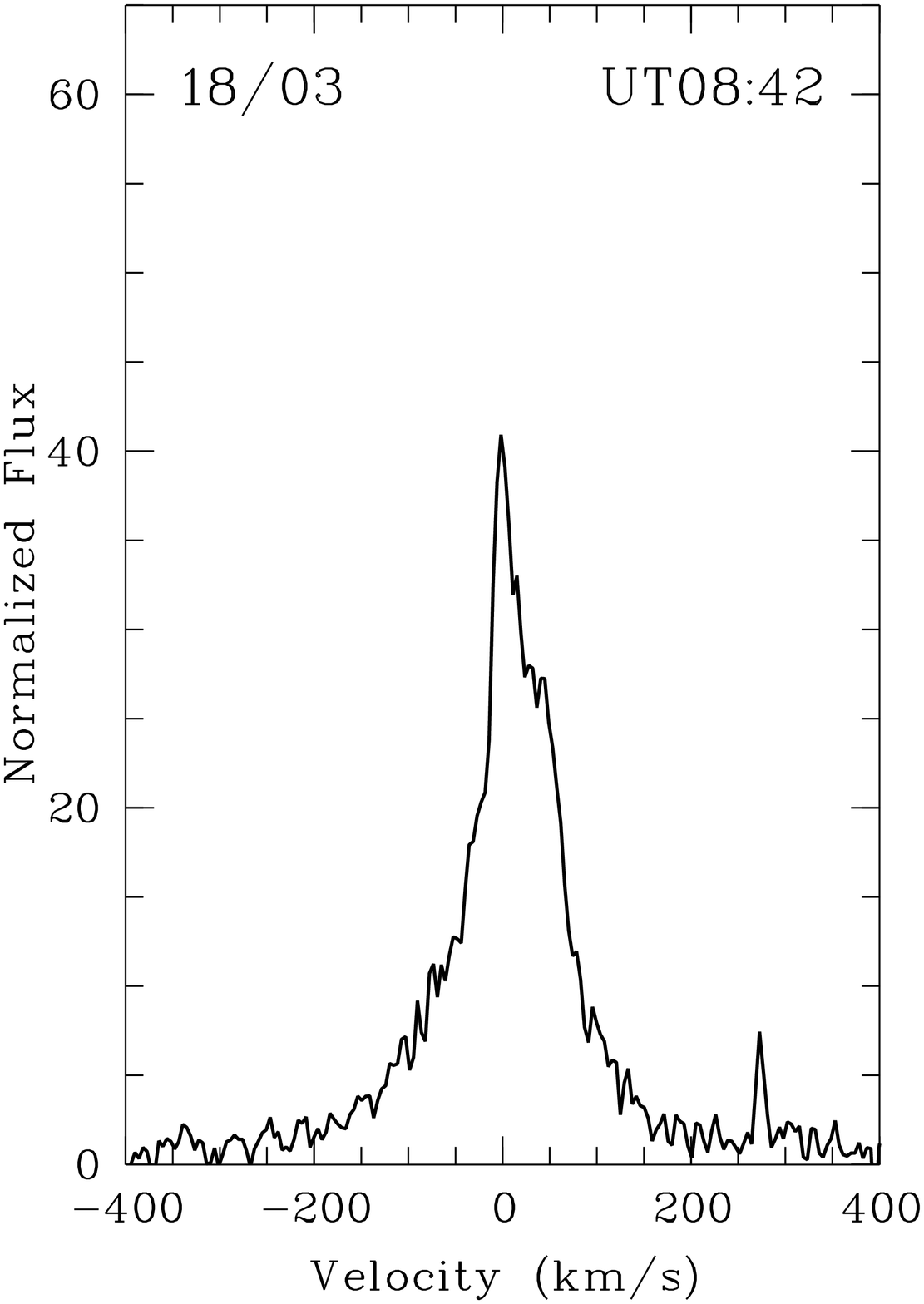}\\
\includegraphics[width=4.0cm]{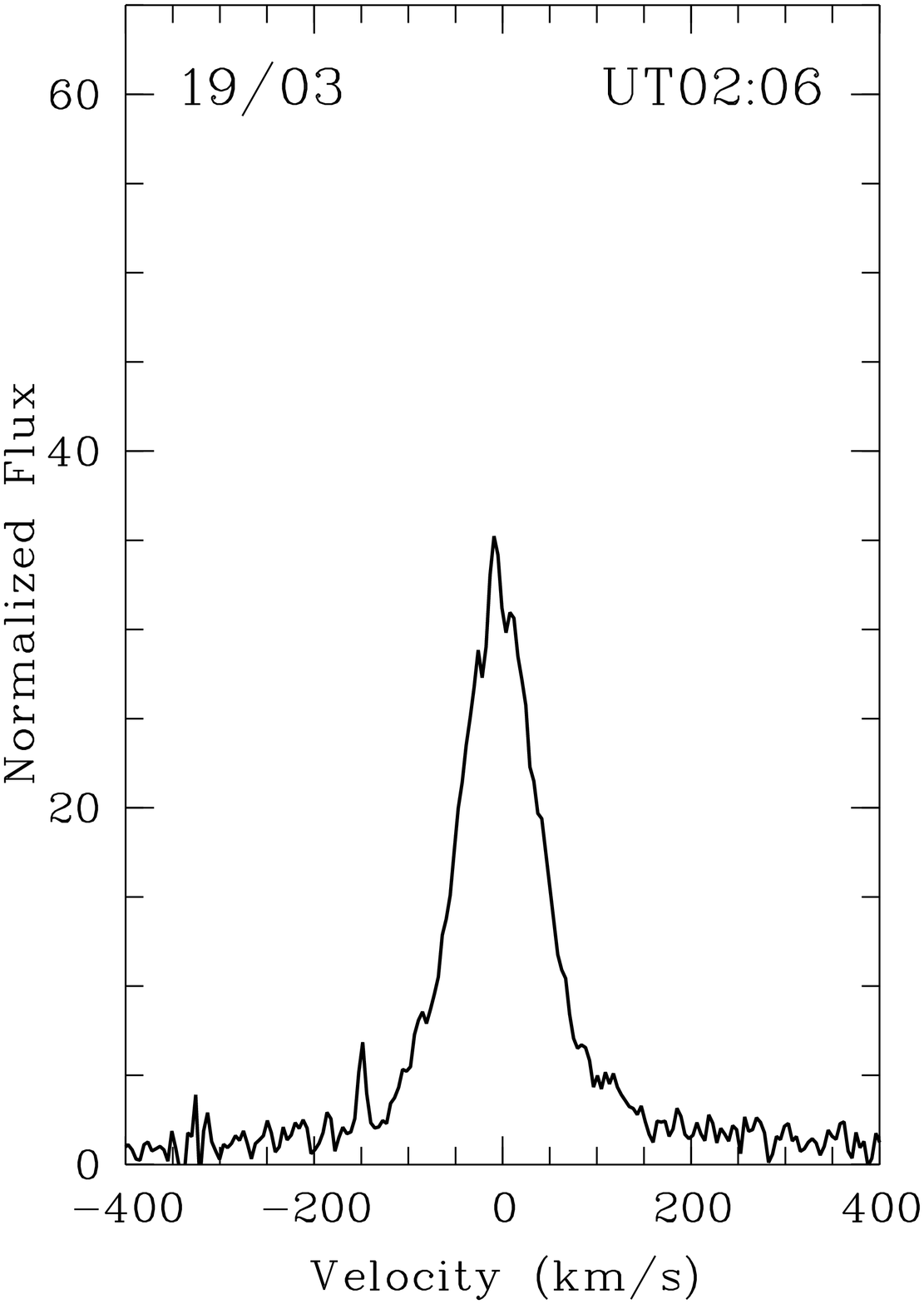}
\includegraphics[width=4.0cm]{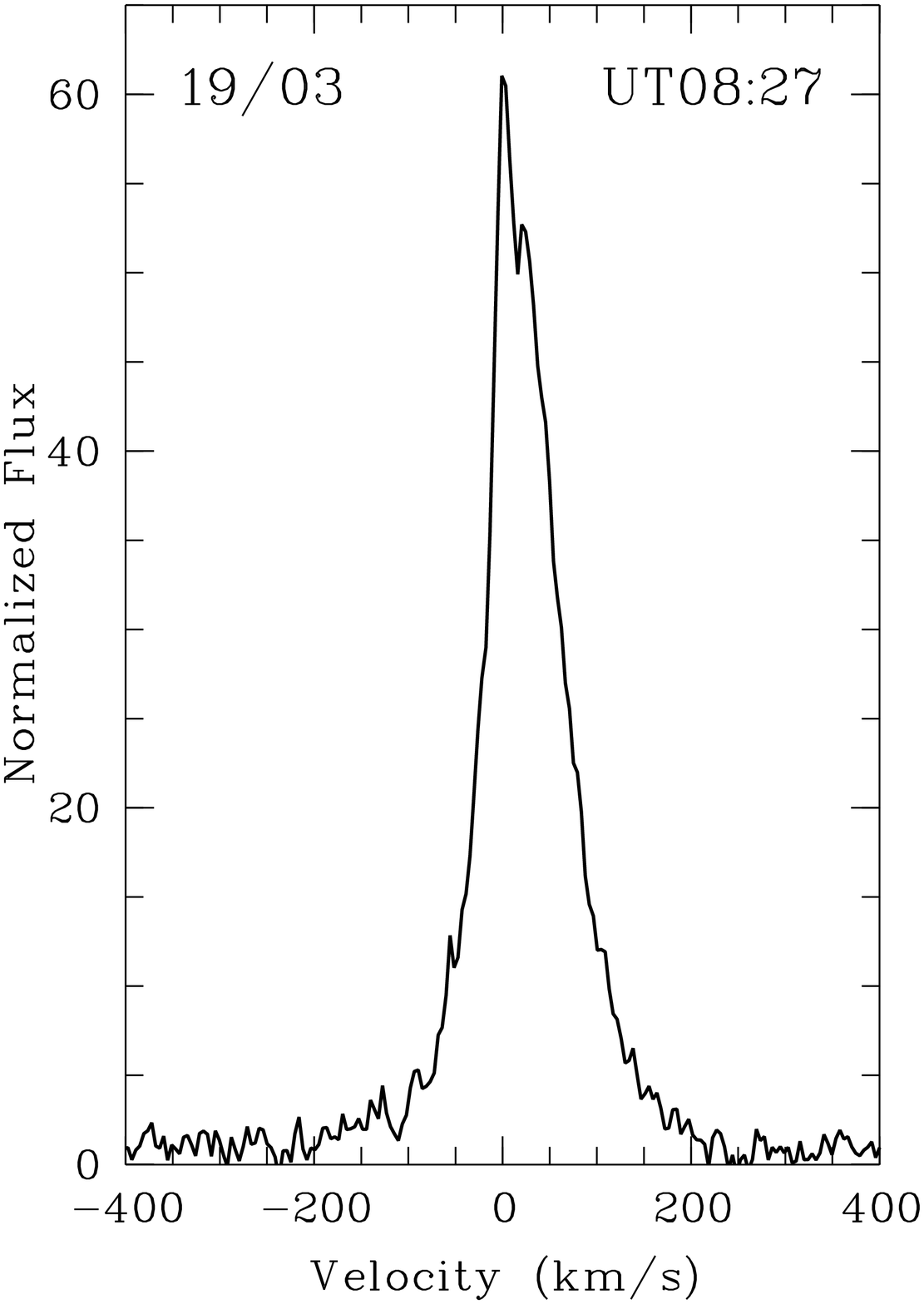}
\includegraphics[width=4.0cm]{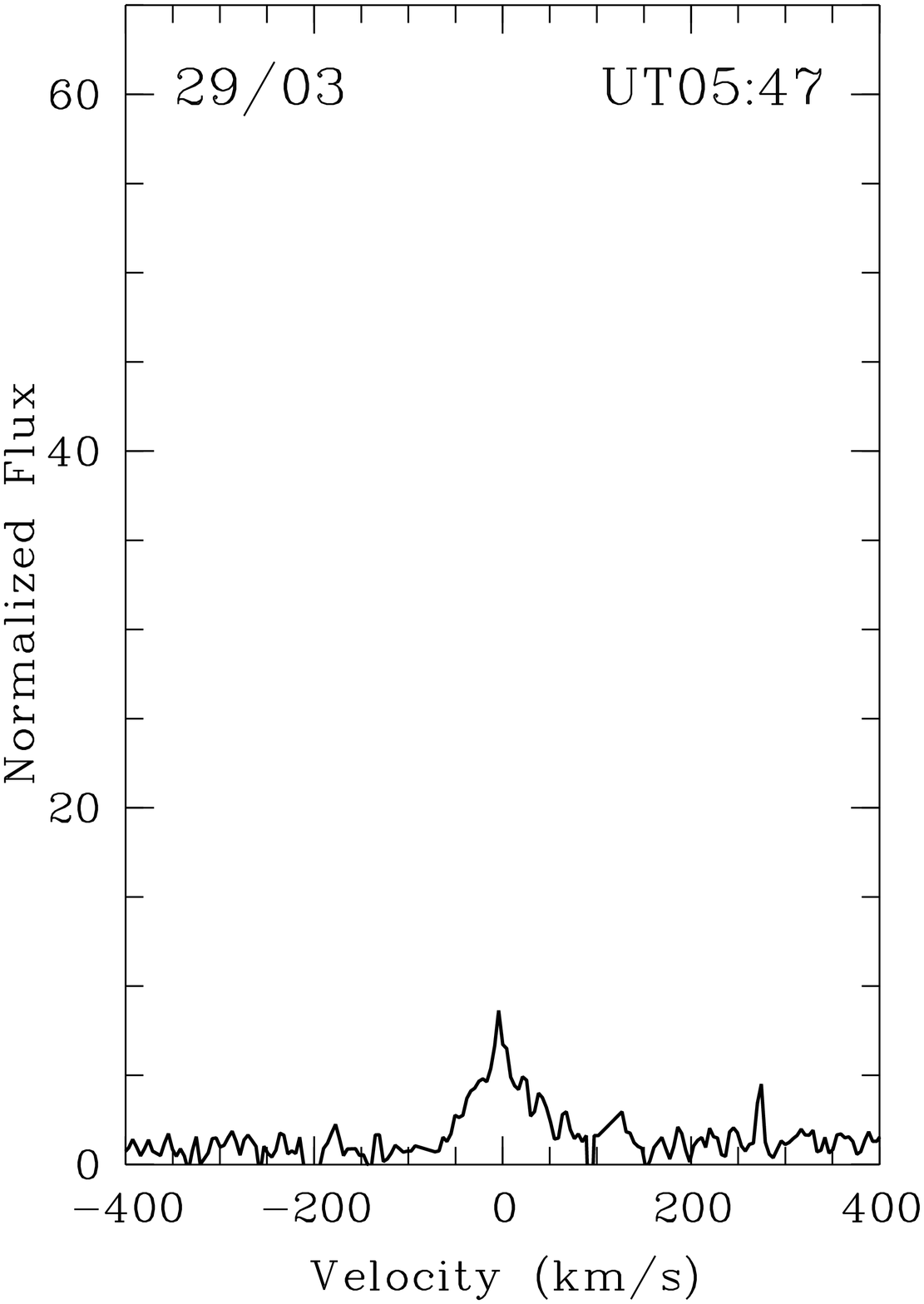}
\caption{H$\alpha$ line profiles for 2M1101. All fluxes are normalized to the continuum. 
Emission features at $\sim 270\,\mathrm{km s^{-1}}$ have their origin in non-perfect 
background subtraction.
\label{ha2m1101}}
\end{center}
\end{figure*}

\begin{figure*}
\begin{center}
\includegraphics[width=4.0cm]{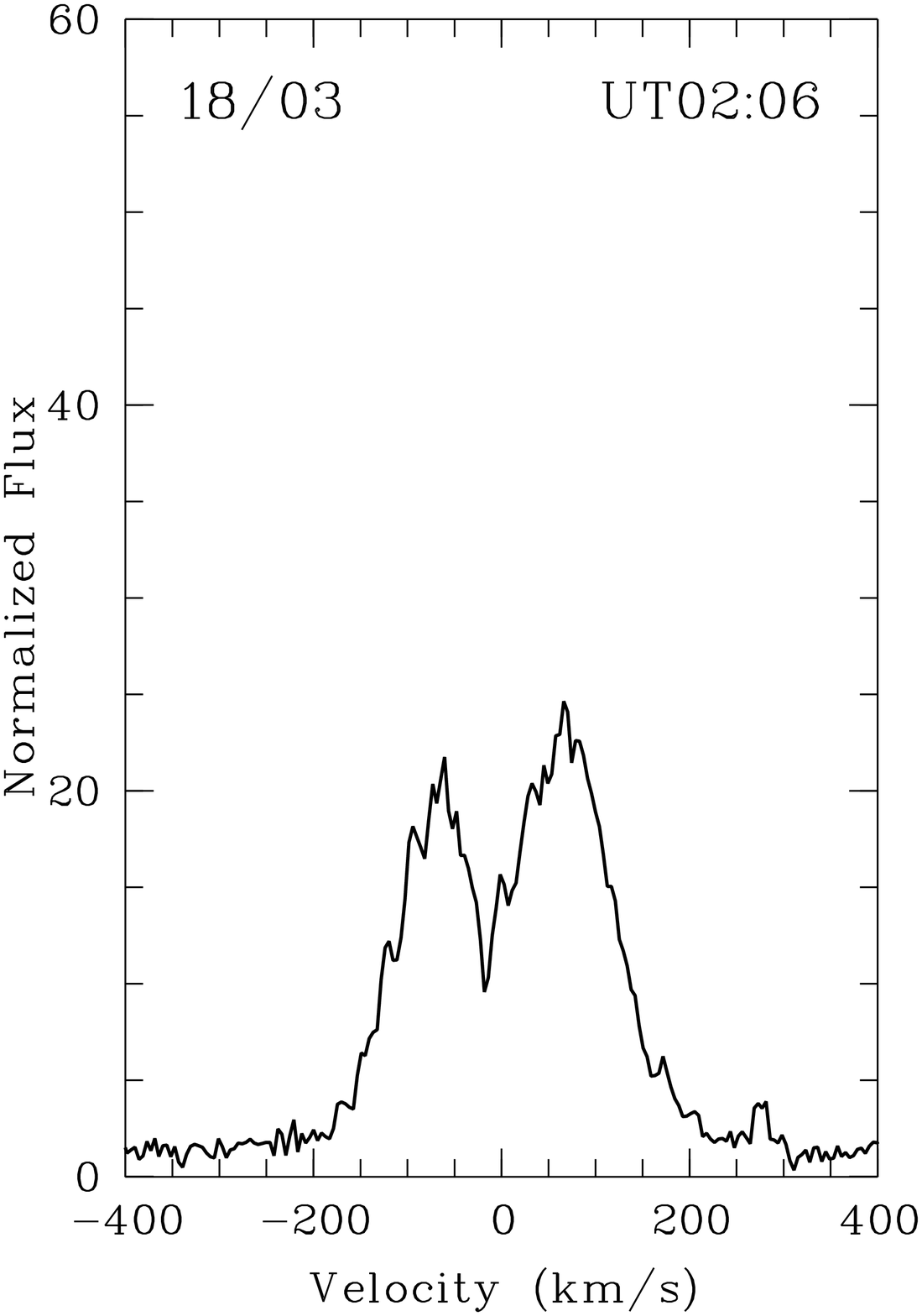}
\includegraphics[width=4.0cm]{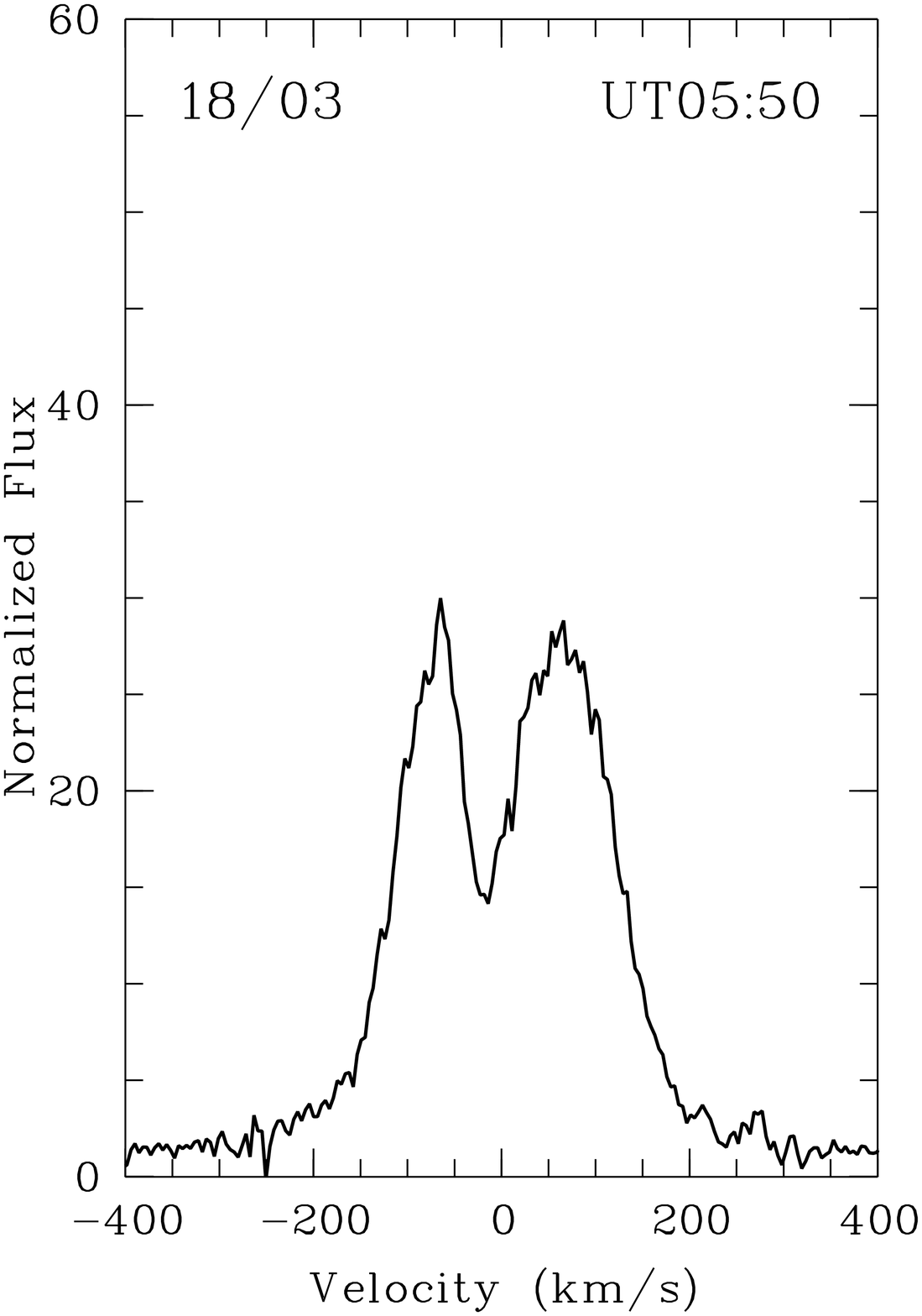}
\includegraphics[width=4.0cm]{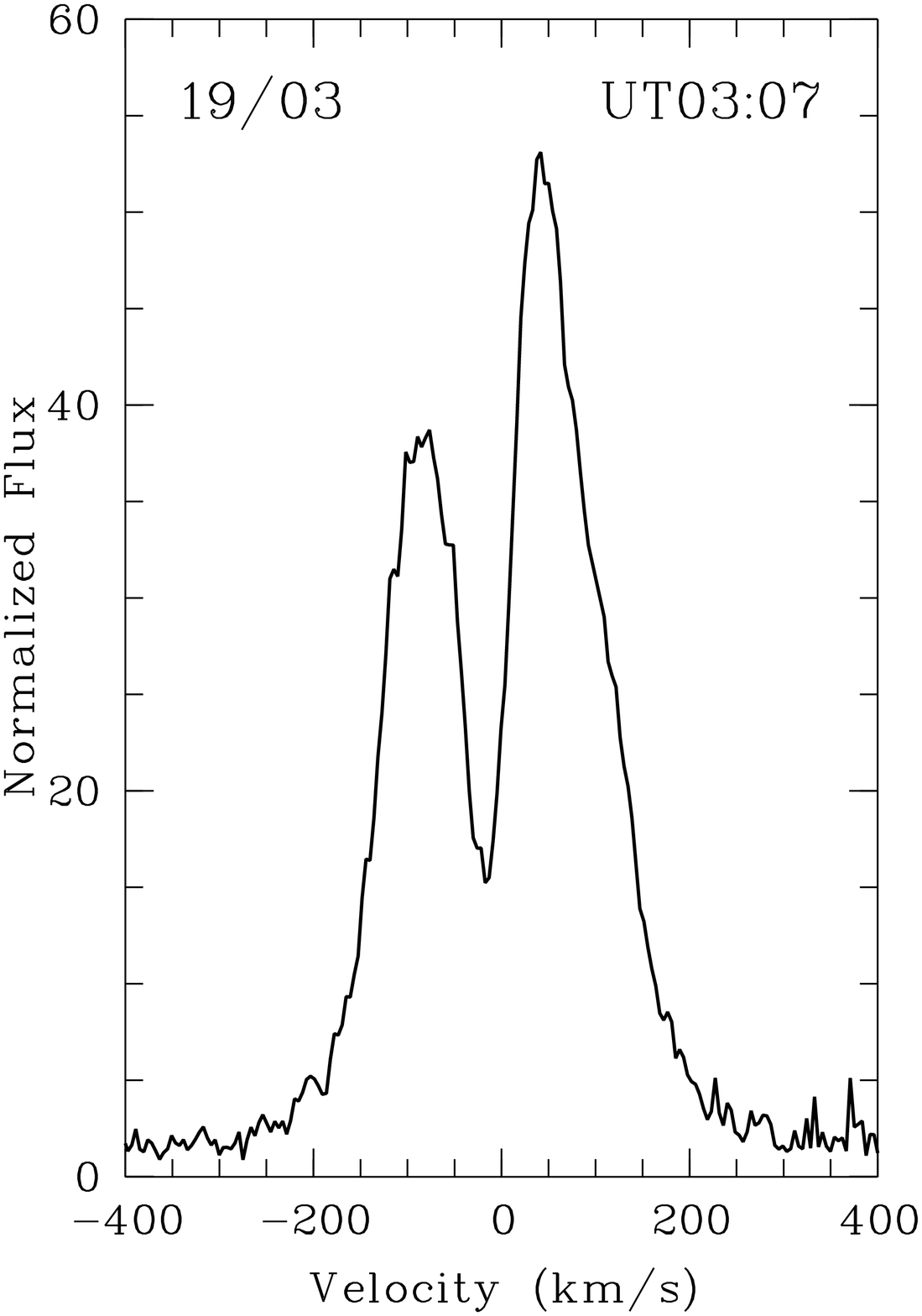}
\includegraphics[width=4.0cm]{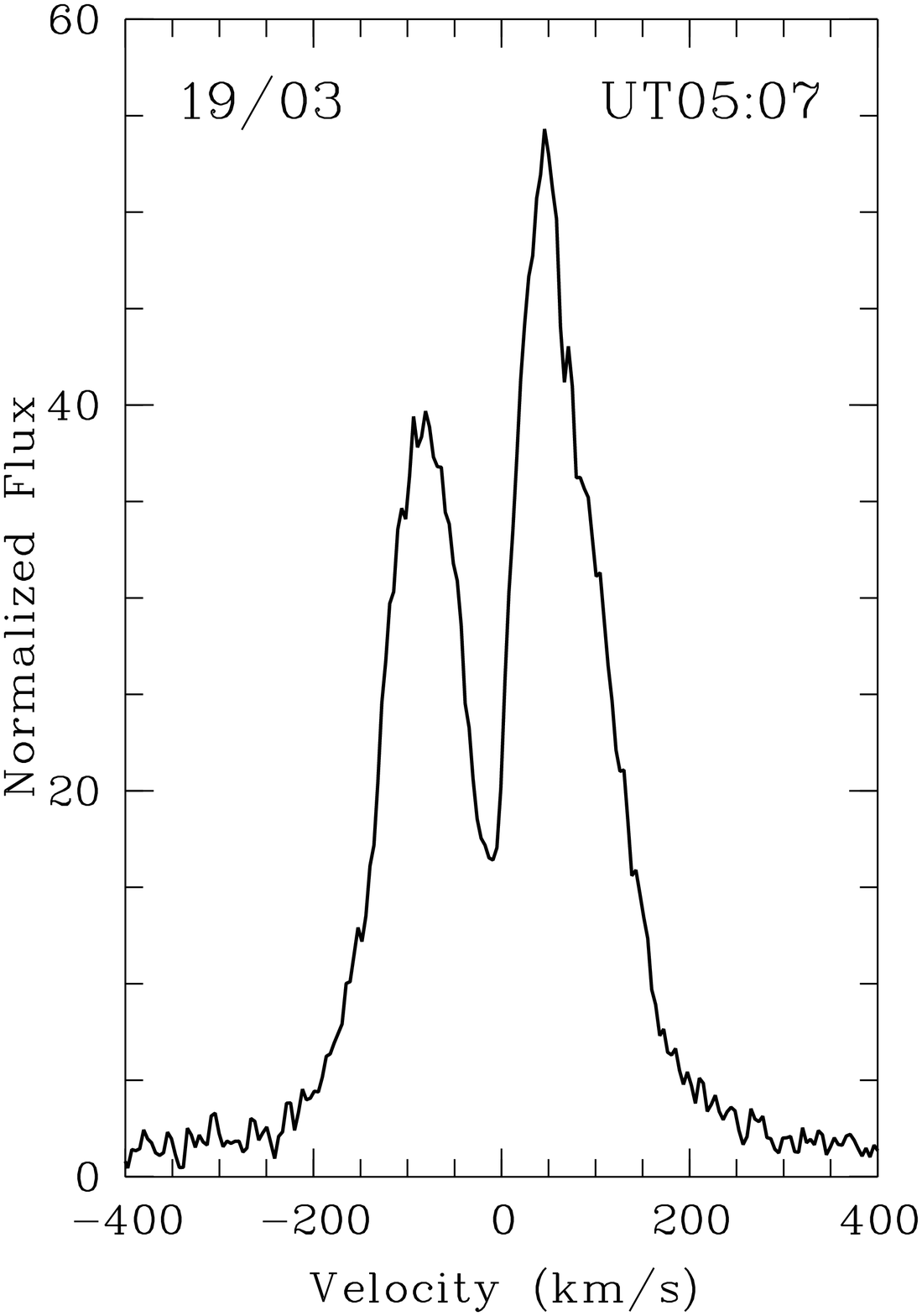}\\
\includegraphics[width=4.0cm]{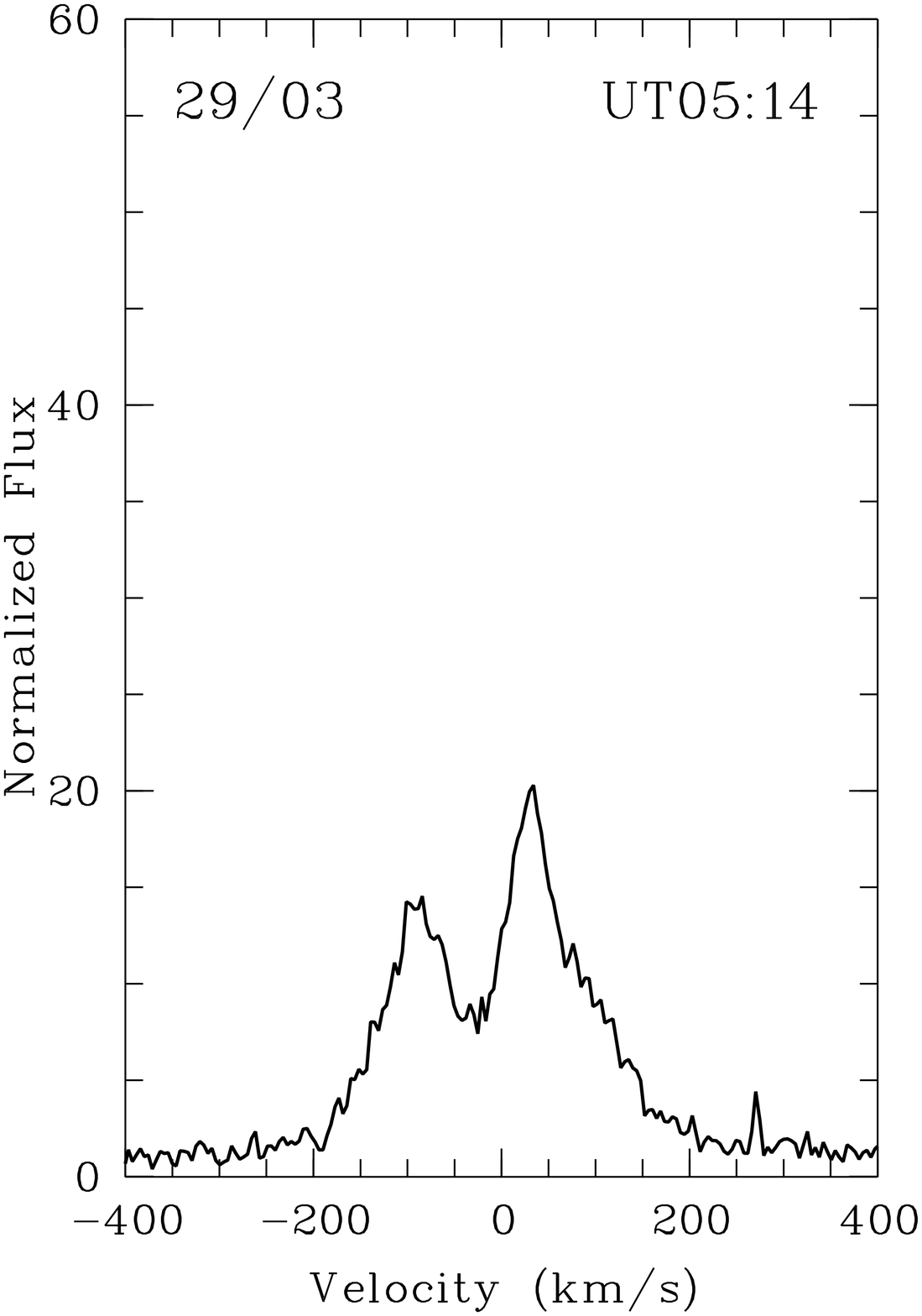}
\caption{H$\alpha$ line profiles for ISO217. All fluxes are normalized to the continuum. 
Emission features at $\sim 270\,\mathrm{km s^{-1}}$ have their origin in non-perfect 
background subtraction.
\label{haiso217}}
\end{center}
\end{figure*}

In Figs. \ref{ha2m1101} and \ref{haiso217} we show the complete H$\alpha$ time series for two 
strongly variable objects, 2M1101 and ISO217. The same plots for 2M1207 are already contained 
in SJB05. 2M1101 exhibits weak, symmetric H$\alpha$ profiles in four spectra, but strongly 
enhanced emission in the remaining three. The object ISO217 has a strongly double-peaked line 
in all exposures, where both the total intensity and the relative strength of the two peaks 
changes significantly. In most cases, however, the red peak is stronger than the blue one.

As already discussed in Sec. \ref{lines}, the 10\% width can be used as a robust measurement
of the accretion rate. Our time series show that the 10\% width is much less sensitive to 
variability compared with the EW: Whereas the H$\alpha$ EWs change by at least 
a factor of two 
in the time series, the 10\% widths are constant with $\pm 15$\% for four out of six targets. 
For two objects, however, the 10\% widths change significantly:
For 2M1207, we reported the strong variability in the 10\% width already in SJB05. We 
found an increase of the linewidth by 32\% between the January and
the March data, i.e. on a timescale of six weeks. In the March data, however, the 10\% width
is variable only to a much smaller extent. Strong linewidth variations are also seen for
2M1101 (see Table. \ref{log}): The 10\% width roughly doubles between spectra 3 and 4, which
are seperated by only 8 hours. In the subsequent exposures, the linewidth gradually 
decreases. The strong increase coincides with the appearance of an asymmetric H$\alpha$ 
profile (see above) and other accretion-related lines in the spectrum (see 
Sec. \ref{survey}), and it is responsible for most of the variability observed for 
this target. We note that the third target which is highly variable in EW and 
profile, ISO217, shows only minor changes in 10\% width.

\begin{figure*}[t]
\begin{center}
\includegraphics[width=5.0cm,angle=-90]{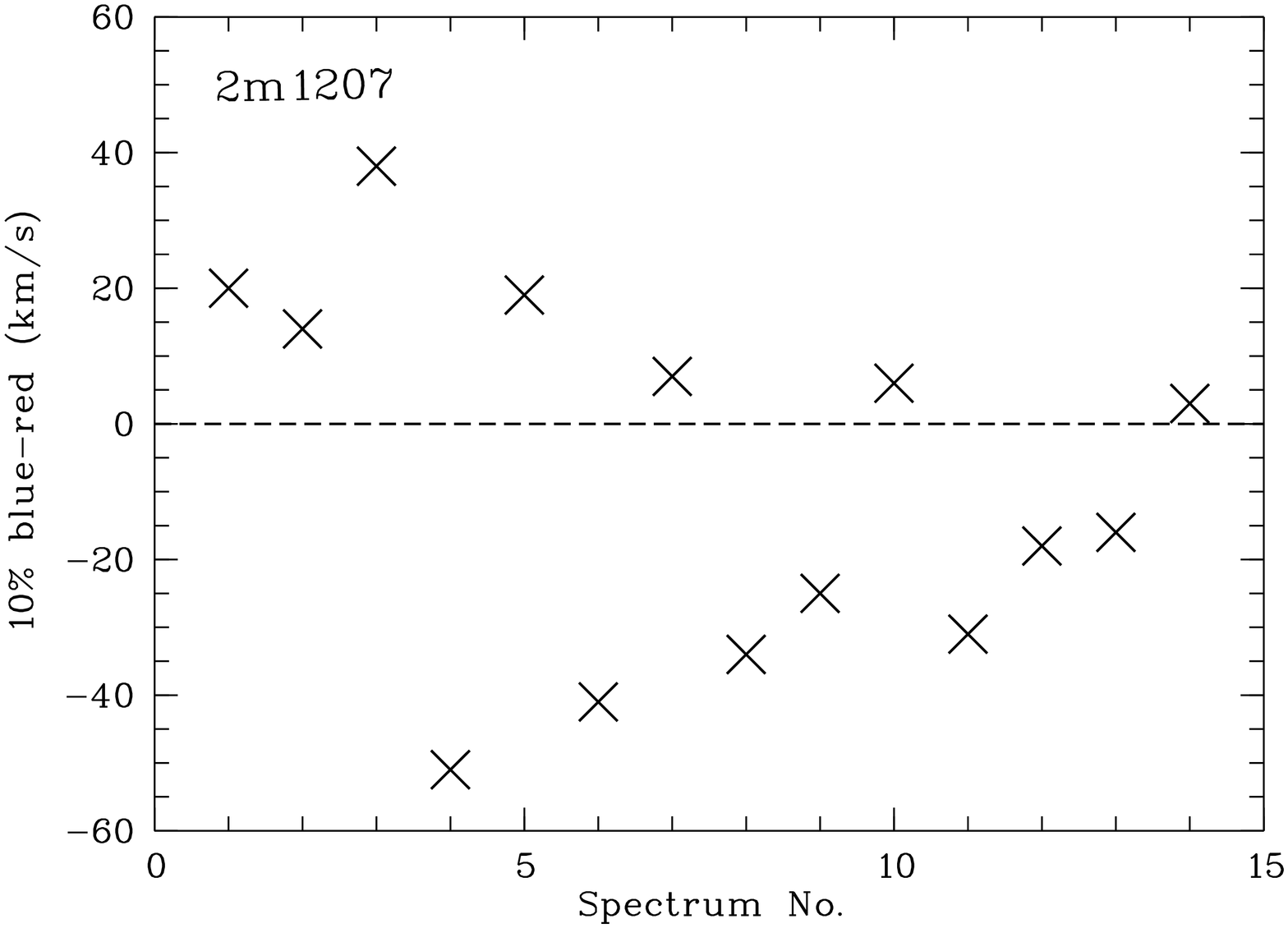}
\includegraphics[width=5.0cm,angle=-90]{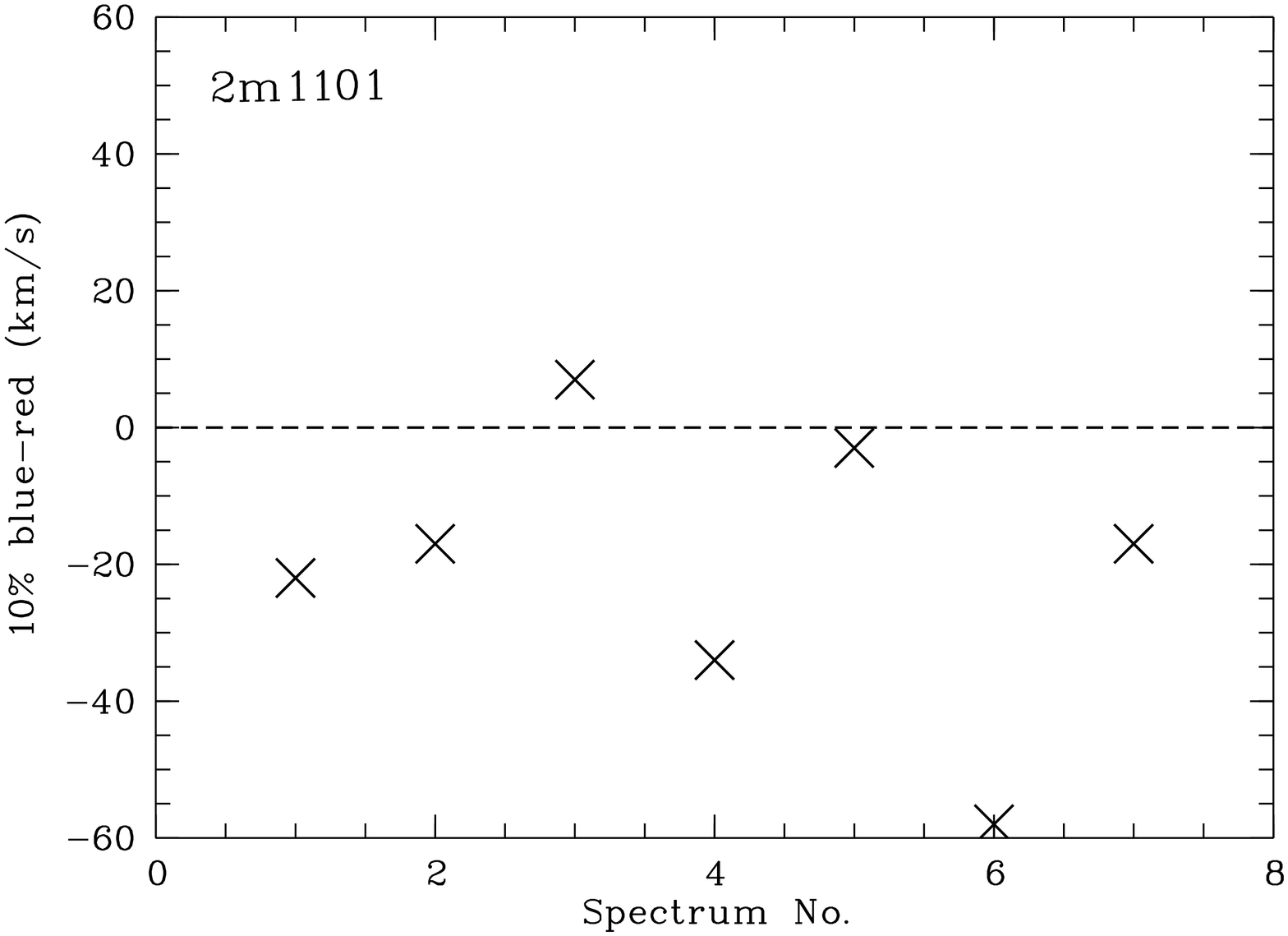}\\
\includegraphics[width=5.0cm,angle=-90]{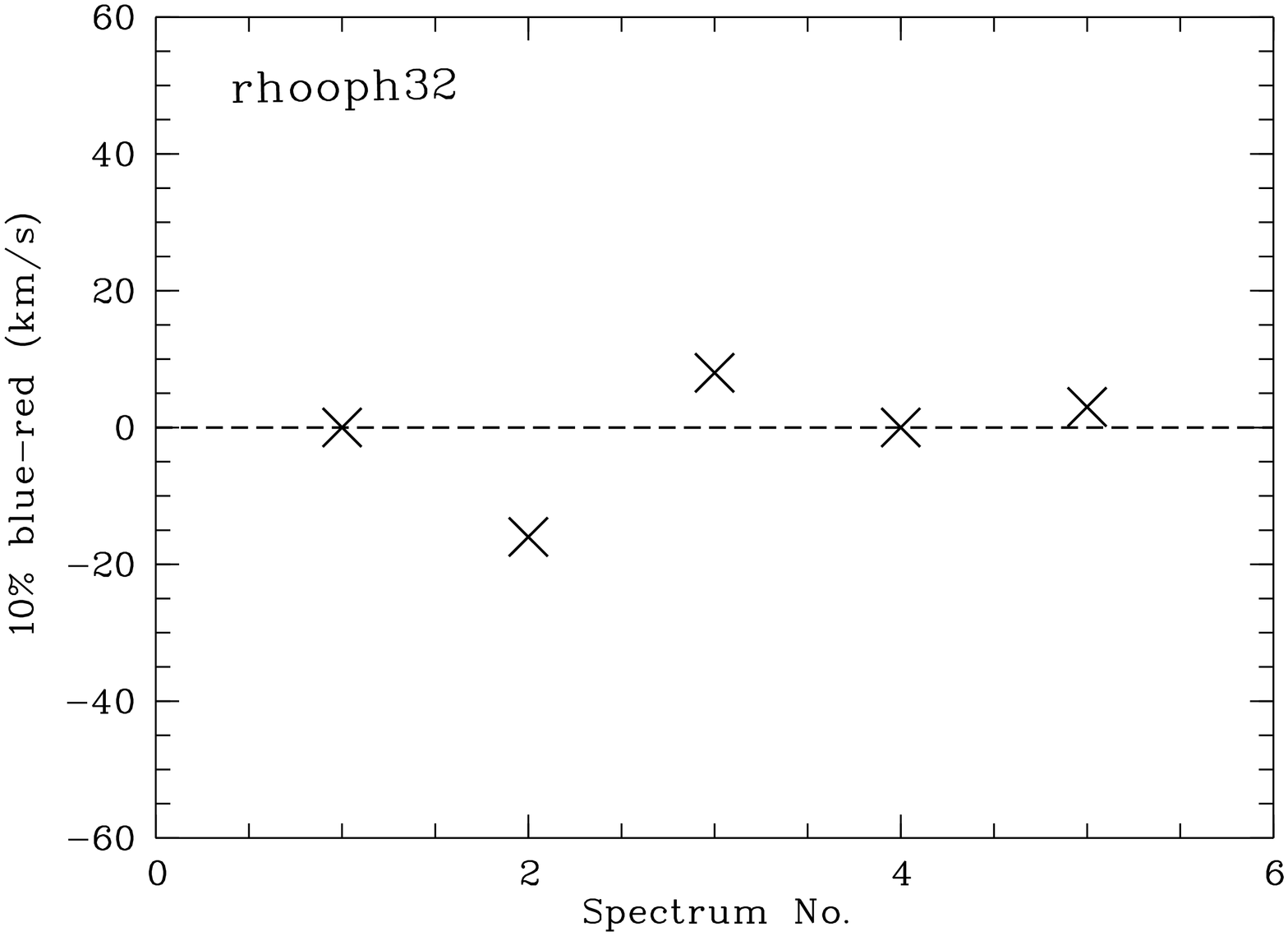}
\includegraphics[width=5.0cm,angle=-90]{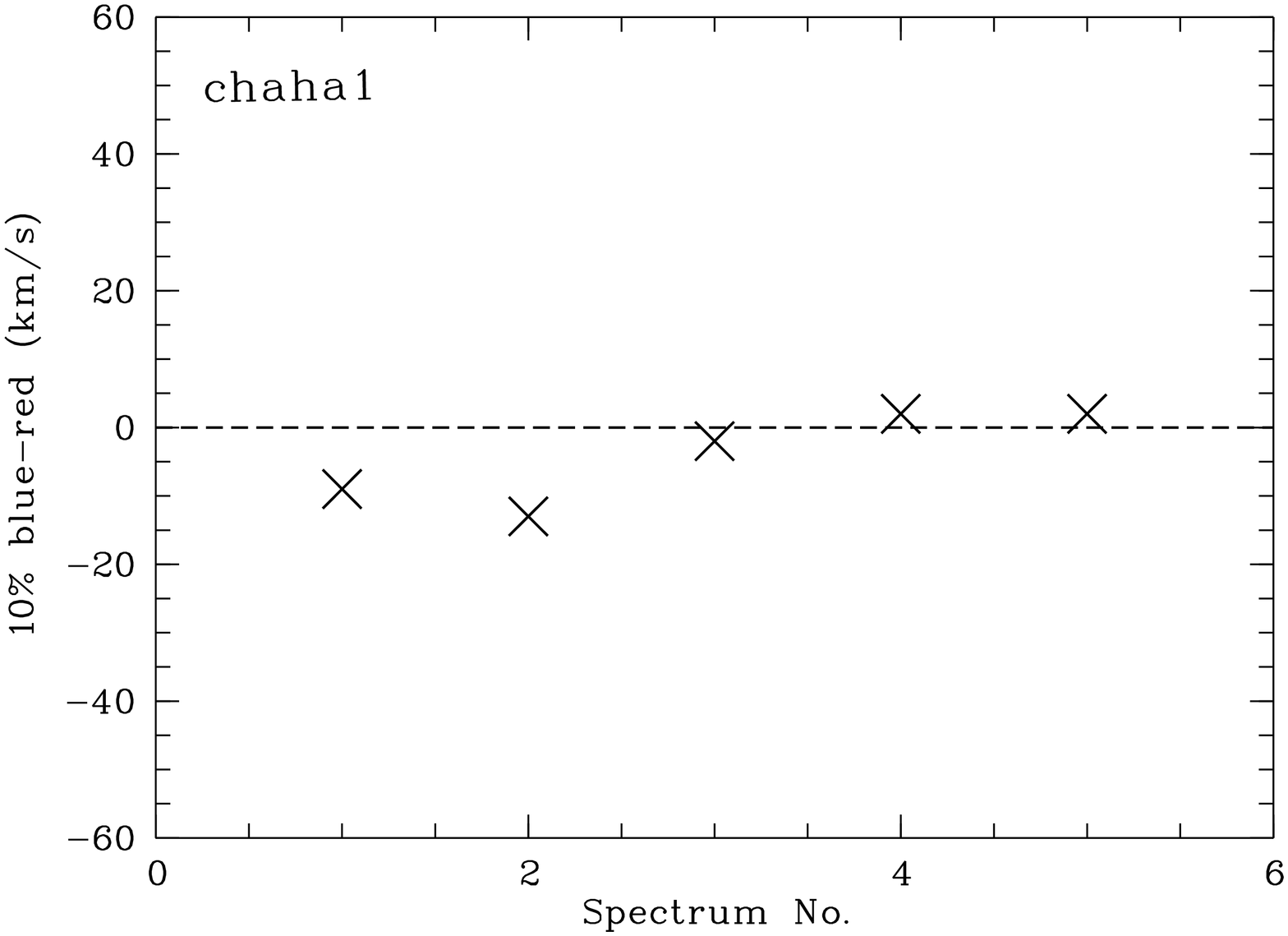}\\
\includegraphics[width=5.0cm,angle=-90]{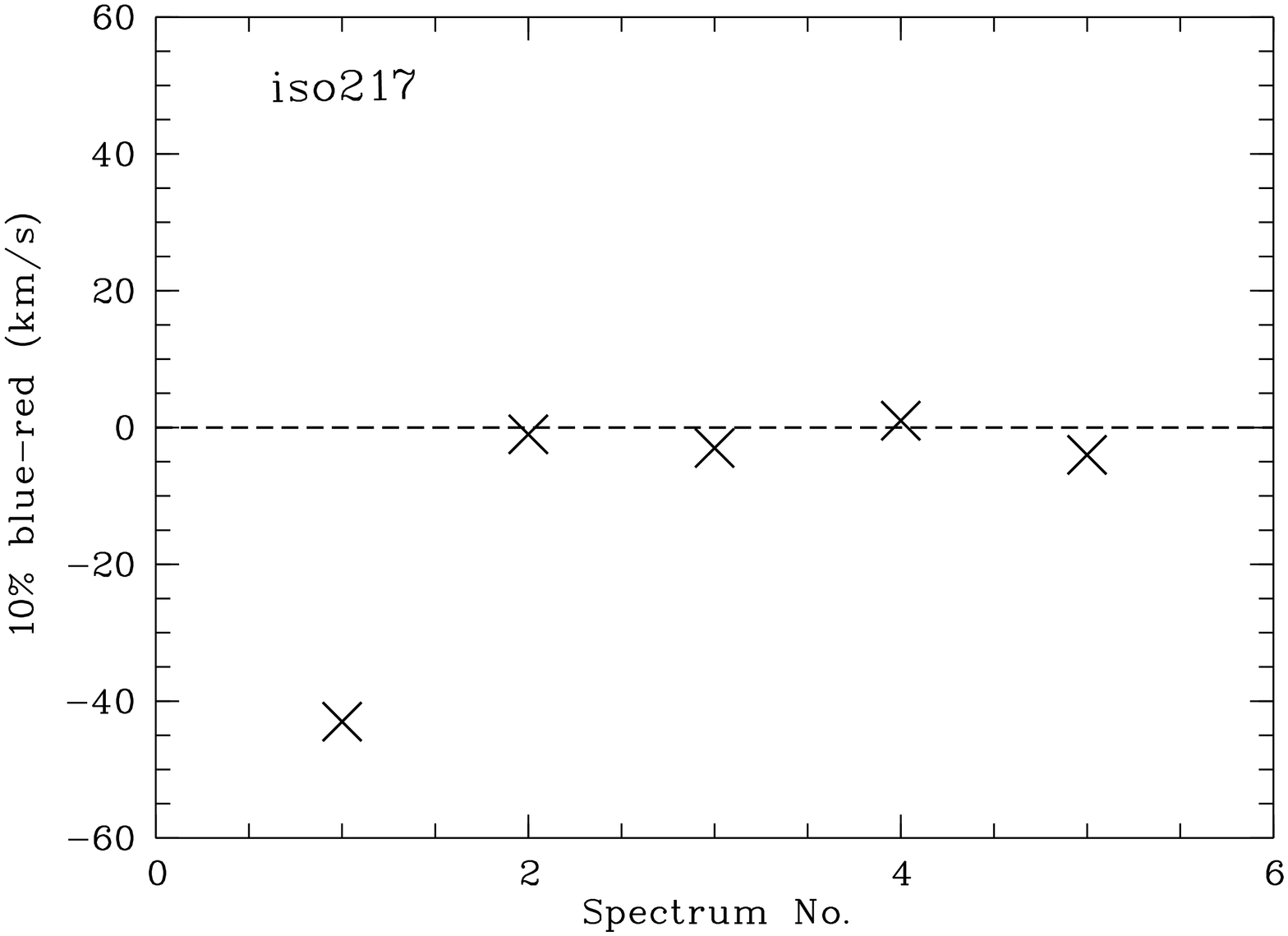}
\includegraphics[width=5.0cm,angle=-90]{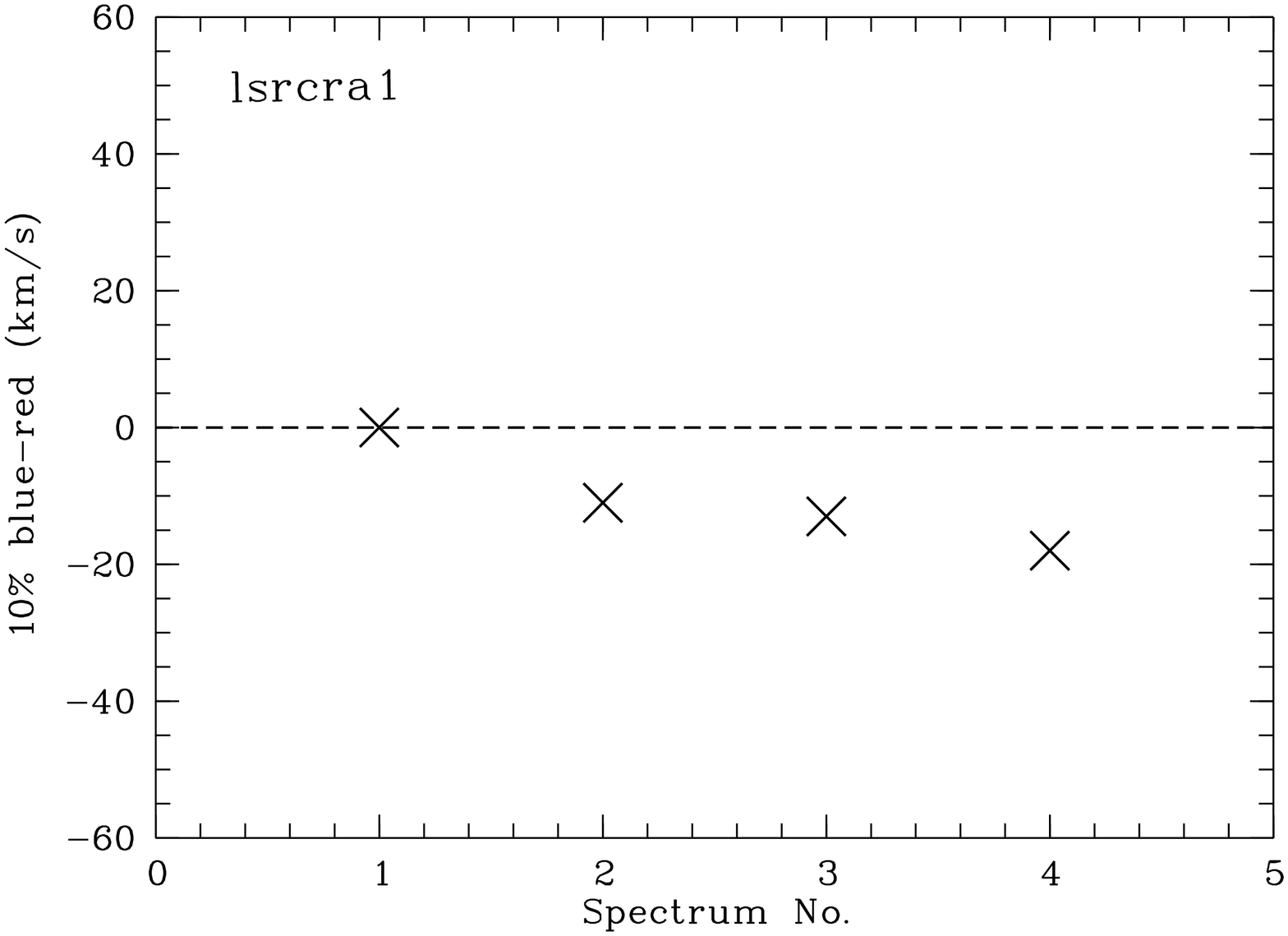}
\caption{Asymmetry in the H$\alpha$ profile, measured as the
difference between blue 10\% width minus red 10\% width. \label{asy}}
\end{center}
\end{figure*}

Another way of quantifying the asymmetry is to measure the 10\% width for blue and red
side of the profile separately. For the three targets without notable asymmetry in
the line profile (ChaH$\alpha$1, $\rho$Oph-32, LS-RCrA 1), the difference between `red' 
and `blue' 10\% width ranges from -20 to $20\,\mathrm{km s^{-1}}$ . The objects 
with asymmetric profiles and high variability in H$\alpha$, however, show strong deviations 
between red and blue linewidths (see. Fig. \ref{asy}). For 2M1207, the line is broader on the 
blue side in spectra with no clear second peak on the red side. As a consequence, 
the difference between red and blue linewidth varies over $\sim 60\,\mathrm{km s^{-1}}$ on 
timescales of a few hours. For 2M1101 the datapoints are mostly negative with an average
of $\sim -20\,\mathrm{km s^{-1}}$, showing that the line is always broader in the red 
compared to the blue. This might be an indication for absorption in the blue part, which 
has been seen before for this object (Muzerolle et al. 2005). As already mentioned, there 
is no clear evidence 
for blue absorption from the profile shape alone. The third strongly variable target, ISO217, 
shows significant asymmetry only in the first spectra, with a difference between red and blue 
linewidth of $\sim -40\,\mathrm{km s^{-1}}$. 

\begin{figure}[t]
\begin{center}
\includegraphics[width=6.0cm,angle=0]{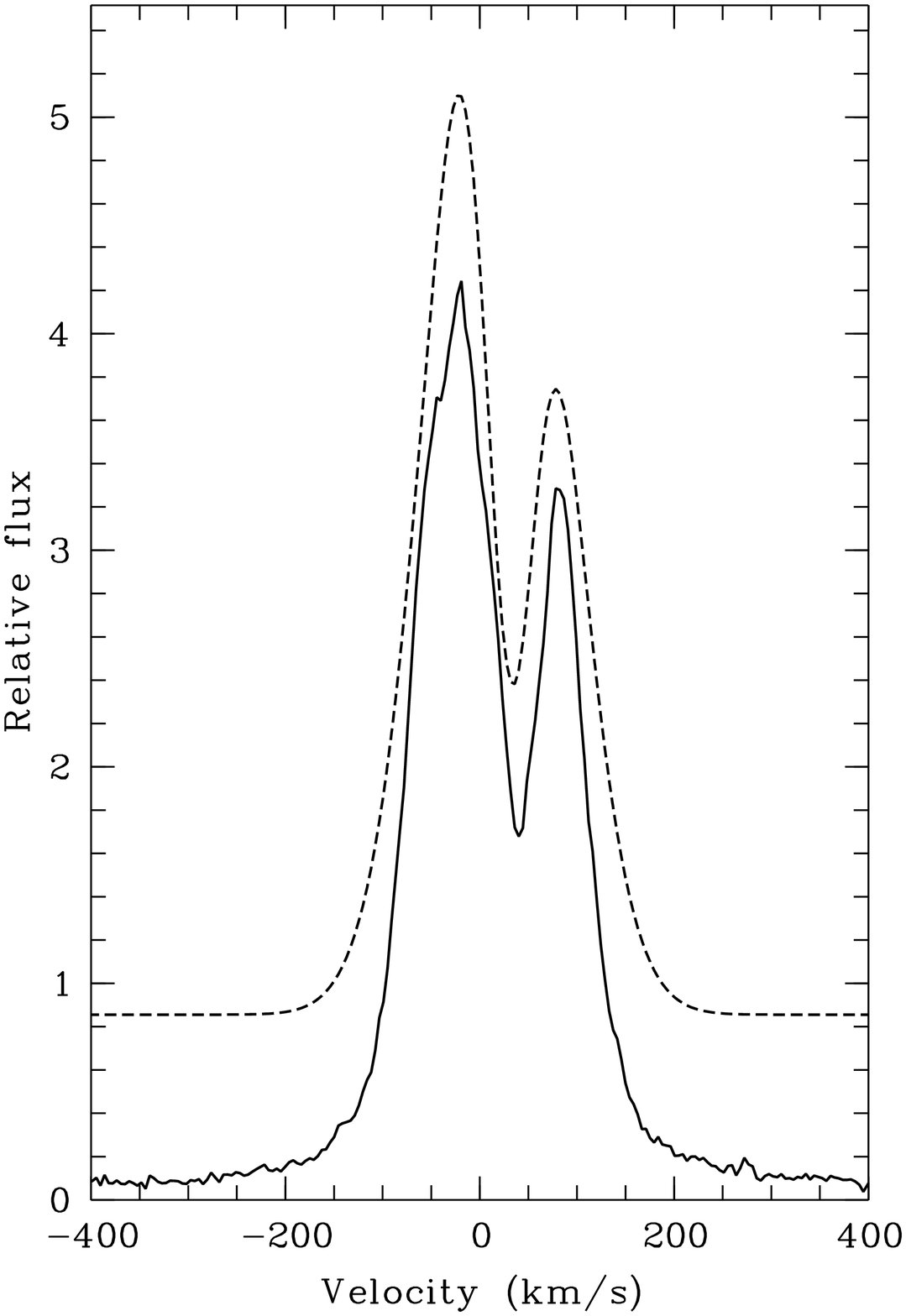}
\includegraphics[width=6.0cm,angle=0]{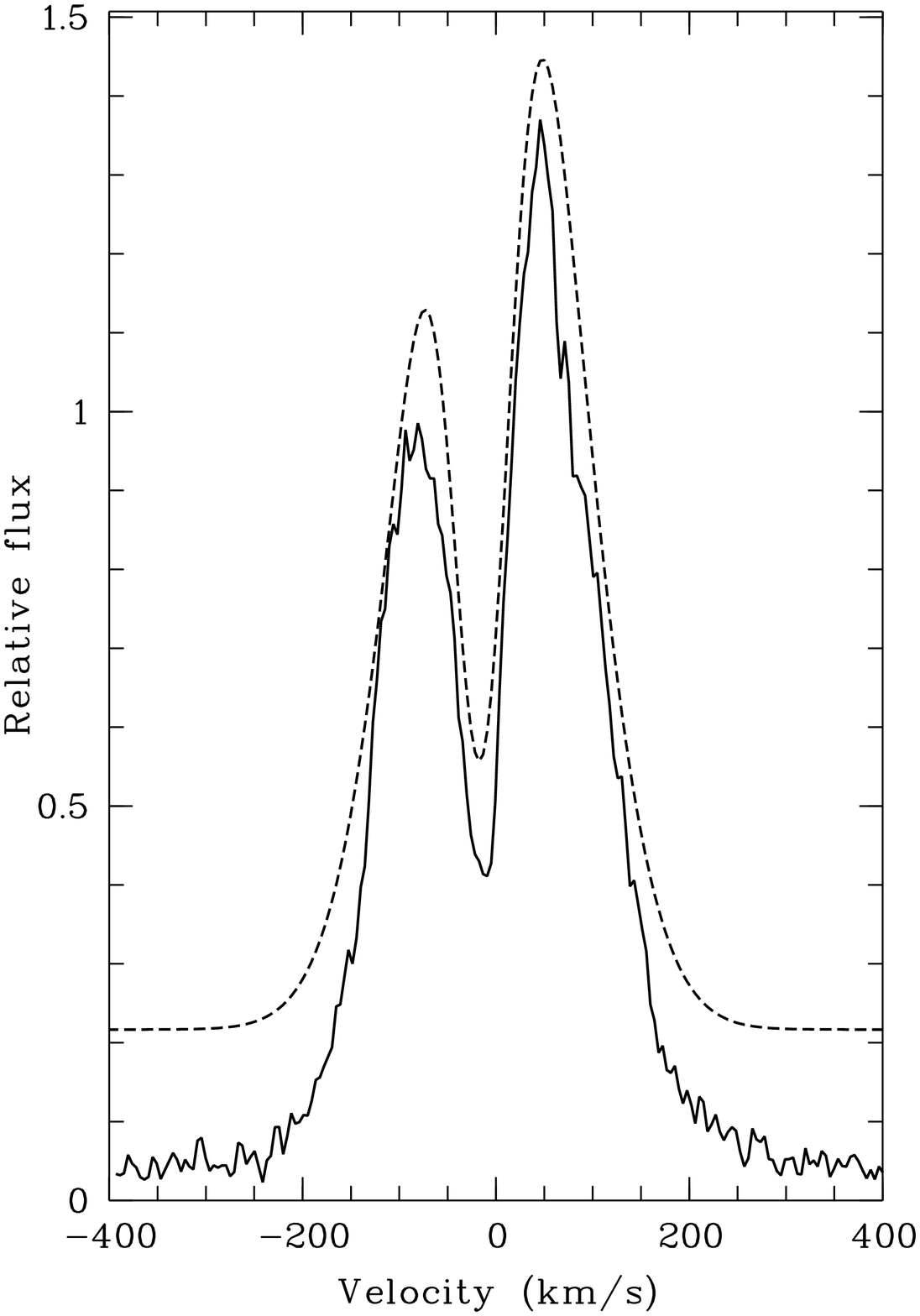}
\caption{Gaussian decomposition for 2M1207 (left panel) and ISO 217 (right panel). 
Shown is the observed (solid) profile from an example spectrum and the fit (dashed), 
which is a coaddition of two Gaussians. The fit curve is shifted by 0.8 (left panel)
and 0.1 (right panel) units for clarity.
\label{gf}}
\end{center}
\end{figure}

Two targets (2M1207 and ISO217) exhibit, at least in some cases, a double-peaked
structure in H$\alpha$ that we interpret as a superposition of a broad emission and an
absorption component. To verify this hypothesis, we decompose the line for all clearly
double-peaked spectra by adding two Gaussians, one for the emission and a second one 
for the absorption component. This decomposition technique is a standard way to 
disentangle the effects of different components in line profiles 
\citep[see, e.g.,][]{jb97,ab02}. The feasability of the method is demonstrated in Fig. 
\ref{gf} for both objects. 

For 2M1207, the best fit closely matches the blue peak and is in reasonably good agreement 
with the red peak. The line wings are more intense than predicted by the Gaussian, but in 
general the fit is acceptable. We are, however, not able to obtain a plausible fit 
if we use two {\it emission} features. In the three double-peaked profiles, both components 
are clearly red-shifted, where the emission is located at $\sim 22\,\mathrm{km s^{-1}}$ and 
the absorption at $\sim 31\,\mathrm{km s^{-1}}$. The linewidths, EWs and fluxes for the 
emission component are higher than those for the absorption component by a factor of 2-3 
in all three spectra.

The decomposition works similarly well for ISO217, where we are also able to reproduce
the line wings with the fit. The broad emission profile and the absorption feature are
both blue-shifted, but in contrast to the emission, which is constant in position within 
our uncertainties (average $\sim -8\,\mathrm{km s^{-1}}$), the absorption feature clearly 
changes its position on timescales of days. It is located at $-4\,\mathrm{km s^{-1}}$ at 
18th March, at $-15\,\mathrm{km s^{-1}}$ at 19th March, and at $-26\,\mathrm{km s^{-1}}$ 
at 26th March. The EW and flux of the emission is higher by a factor 3-4 than the values 
for the absorption.

\section{Emission line survey}
\label{survey}

In the following we give a phenomenological description of the emission line spectrum for each
target seperately. In addition, we check for correlations in the EWs of different lines. In 
general, all spectra show a strong H$\alpha$ line (see Sec. \ref{Halpha}), and most of them 
exhibit other emission lines indicative of either accretion or outflows or both. All objects 
show variability in the emission lines, with the most dramatic changes observed for 2M1207, 
2M1101, and ISO217 as is also the case with H$\alpha$ emission in Sec. \ref{Halpha}.

{\bf 2M1207:} As already discussed in SJB05, the spectrum of 2M1207 is dominated by a 
variety of accretion related emission lines. It shows strong Balmer lines
in some cases down to H10 at 3800\,\AA, both HeI lines, the CaII triplet, and OI. 
We do not see evidence for forbidden line emission in [SII] or [OI] in our spectra. 
The intensity of the H Balmer and HeI lines is strongly variable, the EWs change by a factor
of 5-10. If we separate January and March data, however, the variations in H$\alpha$,
H$\beta$, and HeI5875 are significantly smaller, indicating that most of the
line variability is produced on timescales of a few weeks rather than days. Only HeI6678 
is not strongly enhanced in the March spectra. The CaII triplet lines are weakly detected in 
March, with more or less constant EWs of $\sim 0.6$\AA, and all three lines show roughly the 
same intensity. Similarly, OI is only present in the March spectra.
The EW of all Balmer lines are correlated, where the best correlations are seen between
neighbouring lines in the series, e.g. for H$\alpha$ vs. H$\beta$ and H$\beta$ vs. H$\gamma$. 
HeI5875 is correlated with all Balmer lines, except for two datapoints in March, where this
line shows an enhancement. This 'burst' is also seen in the high-energy Balmer lines, but 
at a weaker level. Our March EW in H$\alpha$ are comparable to the literature values from 
\citet[EW of $\sim 300$\,\AA]{g02}, while the January data are closer to the two other more 
recent literature measurements \citep[EW of 28 and 42\,\AA]{gb04,mjb03}, indicating that the 
changes which we observe for 2M1207 are not unique, and that variability persists on 
timescales of years.

{\bf 2M1101:} Four out of seven spectra for this brown dwarf show weak H$\alpha$ emission
with EW $\sim 15$\,\AA, but in three consecutive exposures on March 18 and 19 the EW is 
increased by a factor of $\sim 6$ (see Fig. \ref{ha2m1101}). This enhancement is also seen in 
H$\beta$ and it coincides with the appearance of H$\gamma$, HeI5875, and CaII lines, which 
are either very faint or not detected when H$\alpha$ is weak. H$\alpha$ and H$\beta$ are clearly
correlated. HeI6678 is only seen in the first spectrum, and neither of our spectra shows
forbidden line or OI emission. We note that the only literature value for the 10\% width 
measured by \citet[$283\,\mathrm{km s^{-1}}$]{mlb05} is higher than any of our measurements, 
whereas their EW is comparable to  our values.

{\bf RhoOph-32:} The Balmer lines and both HeI lines are strong in all exposures, and
their time evolution is comparable: All lines start on a high level, and their intensity
is decreased in the last three datapoints. The best correlations are seen for H$\alpha$ 
vs. HeI5875 and H$\alpha$ vs. H$\beta$. Additionally, we observe weak CaII triplett emission,
which shows a trend comparable to H$\alpha$. We do not detect emission in
[SII], [OI], and OI. \cite{ntm04} published an EW of 48\,\AA, which is at the low end of 
our datapoints. Their 10\% width of $248\,\mathrm{km s^{-1}}$, however, is comparable
with our values.

{\bf ChaH$\alpha$1:} The Balmer lines and the CaII triplet are clearly detected in all 
spectra. 
H$\alpha$ and CaII behave very similarly, with a maximum in the third spectrum, and good 
correlations between the EWs. On the other hand, H$\beta$ and H$\gamma$ have their maximum 
in the fourth spectrum. Both HeI lines can only be seen in the last two exposures, and thus 
behave in a comparable manner to the higher Balmer lines. Clearly, there are two groups of 
lines, where the second group (H$\beta$, H$\gamma$, HeI) appears to react delayed by a 
few hours with respect to the first group (H$\alpha$, CaII). In neither of our spectra do 
we find emission in [SII], [OI], OI. All literature EW and 10\% width measurements for 
H$\alpha$ \citep{cnk00,l04,ntm04,mjb05} are consistent with our results, indicating that 
ChaH$\alpha$1 does not show dramatic changes in the emission lines over timescales of a 
few years.

{\bf ISO217:} The dominant lines are H$\alpha$ and the CaII triplet, which are very strong in 
all exposures; the line ratio of the triplet is always close to 1:1:1. OI is clearly detected 
in all spectra, and H$\beta$ and H$\gamma$ in most of them. Both HeI lines are visible in the 
fourth spectrum, coinciding with the maximum in the higher Balmer lines, and in two out of 
five cases we also detect forbidden line emission in [SII]. The EW in H$\alpha$, CaII, and 
[SII] behave comparably, with a maximum in the third spectrum. On the other hand, the higher 
Balmer lines, HeI, and OI are strongest in the fourth exposure, and are thus apparently delayed 
by $\sim 1$\,h. H$\alpha$ shows a clear correlation with H$\beta$, and weak correlations with 
CaII and [SII]. Our H$\alpha$ EW and 10\% measurements are consistent with literature values 
for this target \citep{l04,mlb05}.

{\bf LS-RCrA\,1:} In our time series we see the Balmer lines, CaII triplet and [SII] in
all spectra, and HeI5875 and OI in most of them. HeI6678, however, is not detected. The 
central CaII triplet line is clearly stronger than the other two, with an average line 
ratio of 1:2:1. CaII and [SII] EWs show the same trend: they increase in the first three 
spectra and drop off in the fourth exposure. In contrast, all other lines show an additional
dip in the second spectrum. Correlations are seen for CaII vs. [SII], H$\alpha$ vs. 
H$\beta$ and H$\alpha$ vs. HeI. Whereas the H$\alpha$ 10\% width is more or less
constant and in agreement with literature values \citep{bmj04}, the EW varies strongly, 
although we do not observe the extreme values of $\sim 360$\,\AA~and $\sim 240$\,\AA~reported 
by \citet{fc01} and \citet{bmj04}.

\section{Discussion of specific targets}
\label{disc}

In the previous two sections we gave a purely phenomenological description of the
emission line variability for our six targets. In this section, we will interpret these
results, aiming to understand the origin of the variability and the line profiles. This
has to be done for each target separately, due to differences in their behavior. 
We would like to emphasize that many of the conclusions in the following
subsections have to be somewhat speculative, since the large number of free 
parameters usually allows more than one possible interpretation for the spectral 
variability.

\subsection{2M1207}

2M1207, which is a likely member of the $\sim 8$\,Myr old TW Hydrae association has 
already been discussed in some detail in SJB05. Therefore, here we wish 
to recall the scenario used in this previous paper to explain the emission line
variability. In SJB05 we discussed the pronounced red-shifted absorption component
in the H$\alpha$ profiles of 2005 March; this feature disappears and re-appears 
on a $\sim$1-day timescale, comparable with the rotation period of 
the object. Since red-shifted absorption indicates that we see cool, absorbing
gas projected against hot material, we interpreted this behaviour as a consequence
of a nearly edge-on view, where the inclination between rotational axis and
line of sight is $>60$\degr. As the object rotates, the accretion funnel moves
through the line of sight. When we see a strong red-shifted absorption feature,
we may be looking into the accretion column. This view is supported by the evolution
of the flux ratios between H$\alpha$ and higher Balmer lines: In spectra with a 
strong red-shifted absorption component in H$\alpha$, the line fluxes in the higher 
Balmer lines tend to be decreased relative to H$\alpha$. Thus, if absorption
is present, we see on average cooler gas than in the symmetric case, where the hot 
spot is unobscured.

Additionally, we found a significant increase in the H$\alpha$ 10\% width between
the January and the March run, indicating an accretion rate change by a factor
of 5-10. We would like to stress that this burst of accretion on timescales of
six weeks is seen not only in H$\alpha$; except for HeI6678, all other emission lines
become significantly stronger in the March spectra. In fact, most of the variability
in the EW is produced on timescales of weeks rather than days. Thus, the strong
profile changes seen on timescales of hours in March do not significantly affect
the global accretion properties of the object. 

The variance profile (see Fig. \ref{avvar}) of 2M1207 clearly shows that most of
the H$\alpha$ variability is produced on the blue side of the profile with a strong
peak at about $-100\,\mathrm{km s^{-1}}$, although there are also peaks
at zero and red-shifted velocities. Since strongly blue-shifted peaks in the variance
profile indicate strong variability in the wind, this could simply mean
that the increase seen in the 10\% width is mainly due to a burst in a hot wind.
On the other hand, although the total 10\% width increases by about $70\,\mathrm{km s^{-1}}$
between January and March, the difference between red and blue part of this value
does not change significantly, as long as no second peak on the red side is visible
(see Fig. \ref{asy}). Therefore, the 10\% width increases both on the red and on the 
blue side. The most likely interpretation is that we observe a change in the accretion 
rate accompanied by a hot wind.

In two spectra in March, some lines (He5875, H$\beta$, H$\gamma$) show a brief
enhancement over a few hours, but H$\alpha$ is not unusually strong. The origin
of this event is unclear, it might be due to a flare in a hot chromospheric region,
or another accretion-related burst.

\subsection{2M1101}

2M1101, a candidate late-M type object in the ChaI star forming region, appears to 
be a non-accreting, weakly active brown dwarf in four out of seven spectra. In the 
remaining three spectra, however, we see a clear emission burst: EW and 10\% width 
in H$\alpha$ strongly increase, and reach levels sufficiently high to classify the 
object as an accretor. From the 10\% width we estimate an accretion rate of about 
$10^{-11}\,M_{\odot} \mathrm{yr^{-1}}$ \citep{ntm04} during this burst. At the same 
time, other lines like CaII, H$\gamma$, HeI5875 appear in the spectra, which are at 
least partly related to the accretion process (see Sec. \ref{lines}). Furthermore, 
the H$\alpha$ profile exhibits a red absorption feature, which is clear evidence for 
infalling material. Thus, based on 10\% width, H$\alpha$ profile, and the appearance 
of additional emission indicators, we argue that this event is related to accretion 
and not, e.g., a flare of chromospheric activity. 

The origin of this behaviour is unclear. It is unlikely that we just see rotationally 
modulated emission, i.e. a strongly asymmetric accretion flow, because the increase 
in the line widths happens within a few hours, whereas the high level is maintained 
for at least one day. A rotational modulation would imply a more periodic behaviour.
To definitely exclude a rotational origin for the observed variability, a $v\sin i$
measurement (which is not possible based on our data, see Sect. \ref{cont}) would
be desirable.

The alternate explanation is a sudden increase of the accretion rate by at least
one order of magnitude. We have seen a similar increase for 2M1207 (SJB05) on timescales
of a few weeks; here we may have the first evidence for strong accretion rate 
changes on timescales of a few hours. Such behavior would be expected if the accretion 
is not continuous, but rather clumpy, with a typical timescale in the range of a few 
days for the discontinuity. Whether or not this applies to 2M1101 could be clarified 
by further time series observations with better sampling.
 
2M1101 has been observed before with high-resolution spectroscopy by \citet{mlb05}. 
Remarkably, H$\alpha$ shows an extreme P Cygni profile with strong absorption on the blue 
side in their spectra from December 2003/January 2004. This has already been seen for some 
T Tauri stars with high mass-loss rates, and the usual interpretation is 
that the H$\alpha$ emission is predominantly formed in the wind. In contrast with this 
literature spectrum, our spectra show only weak evidence for outflowing material as an  
asymmetry in the H$\alpha$ line profile (see Sect. \ref{Halpha}). This implies that both 
the infall and outflow rates are highly variable with time, again underlying the need for 
more detailed follow-up observations.

\subsection{$\rho$Oph-32}

$\rho$Oph-32 has broad emission lines, yet no dramatic line profile and intensity
variations are seen within our time series. The H$\alpha$ width exceeds 
$200\mathrm{km s^{-1}}$ in all spectra, implying accretion rates between $10^{-10}$ and 
$10^{-11}\,M_{\odot} \mathrm{yr^{-1}}$ \citep{ntm04}, and the asymmetry in the profile
as well as the persistent emission in CaII support that $\rho$Oph-32 is an active
accretor, in agreement with previous studies of this object. 

In contrast to the literature spectrum of \citet{ntm04}, we see no forbidden line emission 
and no redshifted absorption in H$\alpha$. Instead, our H$\alpha$ profiles show a weak 
'shoulder' on the blue side, which may be due to a weak blue-shifted absorption feature.
Thus, there is only weak if any evidence for wind in our spectra. We conclude that outflow
and accretion in $\rho$Oph-32 are probably variable on timescales of years, but more or 
less constant over a few weeks.

An appealing interpretation for the lack of variability is a face-on geometry for the 
disk and the central object. In this case, rotation can be excluded as a major cause of 
variability, since we see always the same hemisphere of the object. $\rho$Oph-32 
exhibits strong near/mid-infrared colour excess indicative of a disk 
\citep{pmo00,ntc02}, which also favours a pole-on view because the surface area of the 
disk visible to the observer is maximized when viewed face-on. An alternate explanation 
for the lack of variability would be a more or less symmetric field configuration implying 
axially symmetric accretion.

\subsection{ChaH$\alpha$-1} 

The literature data for ChaH$\alpha$-1 are ambiguous with respect to the accretion status
of this object. \citet{cnk00} and \cite{l04} classified it as weakly accreting based on
H$\alpha$ EW, but the low 10\% widths measured by \citet{ntm04} and \citet{mjb05}
led them to suggest that the H$\alpha$ emission of ChaH$\alpha$-1 is predominantly 
caused by chromospheric activity. Our H$\alpha$ profiles of ChaH$\alpha$-1 indeed resemble 
profiles of chromospherically active stars; it is more or less symmetric, not 
extraordinarily broad and exhibits a small central absorption dip, which is usually 
interpreted as self-absorption in the chromosphere. The 10\% widths are clearly below 
the threshold for ongoing accretion, indicating that accretion probably does not 
contribute significantly to the emission line flux. However, we detect HeI6678 
in emission in one of our spectra with an EW of 1.6\,\AA, clearly too high
to be explained only by activity (see Sect. \ref{lines}). Moreover, the EW in H$\alpha$ 
is very high compared with non-accreting, active objects with similar spectral type 
\citep{bm03}. Thus, we cannot rule out that ChaH$\alpha$-1 is weakly accreting, in 
addition to being chromospherically active. Near- and mid-infrared color excesses 
provide additional support for a disk surrounding this object \citep{pmo00,kg01,jas03}. 

There are at least two possible ways to explain why we observe two different groups of 
emission lines
(see Sec. \ref{survey}). Let us first assume that all emission is produced by chromospheric 
activity. In this scenario, the object may harbor two large active regions in the 
chromosphere with different temperatures. One of these regions is mainly responsible 
for the 'high temperature lines' like HeI and the higher Balmer features. The second 
region is cooler and thus does not contribute to the high-energy features. H$\alpha$
and CaII are generated in both regions. The 'delay' seen in the time evolution of both 
groups of lines can be explained with a phase-difference between the two regions.
A similar scenario is possible for the case where ChaH$\alpha$-1 is weakly
accreting. Here, the object would have a hot accretion flow, which may be
responsible for the emission in HeI, H$\beta$, and H$\gamma$, and an active  
chromospheric spot. If this active region appears a few hours earlier on the visible
hemisphere than the accretion hot spot, we would expect a delay in the high-energy
lines. 

In both scenarios, the variability in the emission lines would be due to rotation.
More evidence for this interpretation comes from Fig. \ref{asy}, where we see a 
hint of a period in the evolution of the asymmetry in H$\alpha$. This is expected, 
because both accretion and activity features should change from blue- to redshifted 
as they move through the line of sight. From the rotational velocity given by 
\citet{jg01}, we derive an upper limit for the rotation period of 3.1\,d (using 
radii from \citet{bca98}). The spectral variability implies a period in the range 
of a few days, consistent with the observed rotational velocity. 

We tentatively conclude that the modest variability of ChaH$\alpha$-1 is due to rotation.
This would additionally explain the large discrepancy in the literature values 
for the H$\alpha$ EW, which were derived from single spectra, because the emission
line intensity would depend strongly on the rotational phase. The interpretation of
rotationally modulated emission lines implies an asymmetry in the chromospheric
active regions and (perhaps also) in the accretion geometry, as well as a 
non-negligible inclination between the rotational axis and the line of sight.

\subsection{ISO217}

This ChaI object at the substellar boundary is known to have intense and strongly 
asymmetric H$\alpha$ emission \citep{l04,mlb05}. Our spectra confirm this finding; they 
show broad and asymmetric H Balmer features, clearly indicative of ongoing accretion.
Forbidden line emission provides evidence for outflowing material as well, as also 
implied by the presence of an H$\alpha$ absorption feature that appears clearly 
blue-shifted in most spectra. Correlations between the EW in [SII], H$\alpha$, and CaII 
show that winds influence H$\alpha$ and CaII somewhat, whereas the `high-energy' lines 
which do not show any connection to [SII] are probably dominated by accretion. 

ISO217 clearly shows variability in both accretion and wind related features. In comparison
to 2M1207 and 2M1101, however, the changes in EW and 10\% width are moderate, i.e. the changes
in the accretion rate are well below an order of magnitude. Given our sparse sampling, 
this does not, of course, exclude the possibility of more dramatic changes in its 
accretion behavior. The variability of ISO217 could be due to rotation, since it occurs 
on timescales of a few hours to days, comparable to typical rotation periods for such 
objects. 

As can be seen in Fig. \ref{avvar}, the variability in H$\alpha$ appears to be 
dominated by the emission wings and thus the high-velocity infalling gas. Additionally,
there is a variation of the relative strenghts of the two peaks, which is probably a 
consequence of a velocity change in the wind-generated absorption component. 
As we have shown in our decomposition procedure, the absorption moves towards the blue 
side within our time series by about $20\,\mathrm{km s^{-1}}$ (Sec. \ref{Halpha}), 
which gradually suppresses the flux on the blue side of the profile, leading to an
asymmetry of the peaks.

Perhaps the most interesting feature in our time series is the short time delay of 
$\sim$1\,h between 
HeI and higher Balmer lines and H$\alpha$, CaII, [SII]. As argued above, the difference
between the two groups of lines is that the second group either exclusively or partly
formed in outflows. Thus, the time delay may be explained by a time-lag between wind
and accretion. One explanation for such behaviour is a coupling between outflow
and infall rates, as already seen in T Tauri stars \citep[e.g. for DF Tau,][]{jb97}. As in 
DF Tau, the wind-related feature in ISO217 increases {\it before} the accretion-related lines.
This is expected in a scenario where the wind originates from somewhat cooler regions at
the inner edge of the disk, whereas the high-energy accretion features are formed very 
close to the surface of the object. In this case, the time-lag would correspond to the 
time needed for the infalling material to travel from the inner disk-region, where the wind 
forms, to the object. Assuming free-fall velocity, the distance between the inner disk 
boundary (in the magnetospheric accretion scenario, the distance at which
the disk is truncated by the magnetosphere) and the surface of the object 
would then be in the range of $5 \cdot 10^5$\,km  or about one object radius, 
a value which is not implausible for substellar objects
\citep{mhc03}.

A time-lag between an observed increase in wind and accretion can also be explained by
rotation, if wind and accretion regions are both asymmetric and spatially offset. Since
the wind is believed to be collimated by the magnetic field, the effect could be enhanced
if the axis of the magnetic field is inclined with respect to the rotational axes, 
introducing an additional asymmetry to the system. 

\subsection{LS-RCrA\,1}

The striking emission line spectrum of this object at the substellar boundary has  
been reported by \citet{fc01} and \citet{bmj04}. It shows all spectroscopic signs of ongoing 
accretion as well as evidence for mass outflow. This is confirmed by our data: The 
H$\alpha$ 10\% width is well above the limit between accretors and non-accretors in all 
spectra, and other indicators of accretion and outflow are also seen. Based on the 
10\% width, we estimate an accretion rate of about $10^{-10}\,M_{\odot} \mathrm{yr^{-1}}$, 
which is at the lower end of the range given in the literature \citep{bmj04}. 

The CaII triplet shows a clear deviation from the 1:1:1 line ratio expected for an 
accretion-dominated origin; furthermore, it is correlated with [SII]. Thus, the
CaII emission is probably at least partly formed in the wind, and not only in the
accretion flow. H$\alpha$, on the other hand, behaves in the same manner as other 
accretion-related features, and it shows no evidence for blue-shifted absorption, 
indicating that the line originates primarily in the accretion flow. 

LS-RCrA\,1 is one of the less variable targets in our sample; the H$\alpha$ line is 
more or less symmetric, and does not show significant variability in the profile shape 
or the 10\% width. This may indicate that the system is either seen close to face-on 
or that the accretion flow is fairly symmetric; both scenarios would avoid rotational 
modulation. Based on our time series, we favour the latter explanation, because we
see strong forbidden line emission, but no significant wind signature in H$\alpha$.
which may imply that the H$\alpha$ and forbidden line emission regions are spatially
offset. With a face-on geometry, it is more likely that the [SII] and H$\alpha$ emitting 
regions overlap in the line of sight, producing a wind signature in H$\alpha$. On the 
other hand, with the disk seen close to edge-on, it is much easier to imagine that the 
outflow is seen against the background sky and not against the accretion flow and the 
object itself, and is thus spatially offset from the H$\alpha$ emitting regions, which 
would explain the lack of blue-shifted absorption in H$\alpha$. 

The published studies are inconclusive about the disk orientation of LS-RCrA\,1. On the
one hand, there is some evidence in the literature for an edge-on geometry for LS-RCrA\,1, 
because a) the object shows no colour excess in the near-infrared although it is accreting, 
b) it appears to be underluminous, as would be expected if the object is obscured by the 
disk, and c) the outflow signature is very prominent \citep[see][]{fc01,bmj04}. However, 
the forbidden 
line emission is asymmetric and blue-shifted, indicating that the receding jet may 
be obscured by the disk, which would argue against an edge-on scenario \citep{fc05}. 
Thus, based on currently available data, it is not possible to distinguish reliably 
between face-on and edge-on geometries. 

\section{General implications}
\label{general}

Although our six targets are rather diverse in terms of their accretion properties,
some general conclusions are possible. We presented in this paper the first 
dedicated spectroscopic variability study for accreting brown dwarfs, albeit 
with sparse sampling for most objects. Perhaps the most important result from this 
program is that accreting brown dwarfs, similar to T Tauri stars, are strongly
variable, both in broad band photometric time series \citep{se04,se05} and in 
their emission line spectrum. Variability was originally the main observational 
feature of T Tauri stars, leading to a systematic study of these objects. It is 
therefore worth confirming that variability persists down to sub-stellar masses.
This completes the recently emerged picture of young brown dwarfs as very low
mass T Tauri-like objects, which share many properties with accreting stars.

The broad emission lines of accreting BDs show a large variety of shapes; we 
observe profiles ranging from double-peaked to more or less symmetric, as is the case 
for more massive stars. The intensity and profile shapes 
are variable on timescales ranging from hours to years, again in total agreement 
with CTTS. One implication from this finding is a cautionary note: since
accretion-related properties can be highly variable, investigations based
on one single spectrum or datapoint are not particularly reliable. To study
accretion properties, we either have to work with large samples of objects
(to average over the variability) or with time series. 

One particularly important recent finding, which may be strongly influenced by
variability, is the relationship between object mass and accretion rate. \citet{wb03}
pointed out a positive correlation between $M$ and $\dot{M}$, which 
has been specified more recently as $\dot{M} \propto M^2$ \citep{ntm04,mjb05,mlb05}. This
relationship has been established using a fairly large number of objects, and in all 
samples 
the accretion rates in a given mass bin scatter by about one order of magnitude. In fact,
it is only possible to see the accretion rate vs. mass correlation when sub-stellar
objects are included, and thus the object masses cover more than one order of
magnitude. (Otherwise the scatter is too large.) The derived relationship
has been interpreted recently as a direct consequence of Bondi-Hoyle accretion,
implying that the accretion process in the inner disk is governed by the large-scale
environment of the parent cloud \citep{pkn05}. Although this approach is able to
reproduce the observed correlation, the scatter around the correlation remains 
unexplained.

Variability provides a straightforward solution for this problem. Two out of six
targets (2M1207 and 2M1101) in our sample show accretion rate variations by about 
one order of magnitude. Since we intentionally selected objects where we expected
to see strong changes, these may be extreme cases of accretion rate variability.
On the other hand, given our sparse sampling, our variability estimates should 
be considered only as lower limits; we cannot exclude that our targets show
larger accretion rate changes on longer timescales. Thus, we conclude that
brown dwarfs can in some cases show accretion rate variations of at least 
one order of magnitude.
 
In the Bondi-Hoyle scenario, all objects with a given mass should have the same
(average) accretion rate. Due to variability, the accretion rates based on single 
spectra will scatter by about one order of magnitude around this average value. 
Additional noise may be introduced by geometry and obscuration effects. Thus, 
the scatter observed in the mass-accretion rate relationship is not surprising
-- and it demonstrates that variability studies are an important complement to
understand the accretion behaviour of BDs. 

Variability information can be used to constrain the nature of accretion and winds/outflows 
in sub-stellar objects, in particular to probe the magnetospheric accretion scenario,
which allows for asymmetric accretion and winds, in contrast to spherical infall.
In general, there are two ways to produce variability in the emission lines: 
either the accretion and/or the wind is not steady, which may occur in a scenario 
of stochastic, clumpy flows, and does not necessarily require magnetic funneling, or the 
magnetic field, which funnels the flow, is highly asymmetric, leading to rotationally 
modulated accretion and wind. It is possible to distinguish between the two scenarios by 
from the shape of emission lines -- if the features look asymmetric 
and the profiles change with time, we would expect an asymmetric field geometry. That, 
in turn, would suggest non-spherical accretion, and thus provide indirect evidence for 
magnetospheric accretion. We note that this argument is not reversible: the absence 
of emission line variability or profile asymmetry does not rule out asymmetric accretion.

Among our targets, at least two show strong emission line variability, both in intensity
and profile shape, {\it and} an asymmetric H$\alpha$ feature (2M1207 and ISO217). A third one 
(2M1101) exhibits an asymmetric profile in parts of the time series and a strong accretion 
burst, but as argued in Sec. \ref{disc} the timescales of this event are barely compatible 
with rotational modulation. Thus, at least for one third of our objects, an asymmetric flow 
geometry is required to explain the emission line behaviour, providing a strong case for
magnetic funneling of accretion and winds.

Based on the available data, we can obtain an estimate of the 
magnetic field strengths necessary for magnetospheric accretion in sub-stellar objects, 
using the relationship given by \citet{h02} (Equ. 1), and scaling it from the stellar
regime to the properties of BDs. As as cautious note, we would like to add that this 
should only be considered as an order of magnitude estimate, mainly because of the limitations 
of the equation in \citet{h02}. For example, this relationship assumes a dipolar field 
geometry, whereas there is clear observational evidence for more complex field structure in 
T Tauri stars \citep[e.g.][]{jg02}. Given the absence of magnetic field measurements for
brown dwarfs, a first estimate of the field strength -- albeit crude -- is still useful.

We approximate the mass of a typical BD to be about 
1/20 of a solar mass and the radius to be about 1/3 (at 1\,Myr). Addtionally, 
we scale the accretion rate following $\dot{M} \propto M^2$, and assume that the radius 
at which the disk is truncated by the magnetic field is (in units of the object radius)
more or less constant with mass. The latter assumption is in line with the results
from profile modeling by \citet{mhc03}, where reasonable agreement between observed
and modeled profiles can be achieved without major mass dependency in the (relative) 
size of the magnetosphere. With these prerequisites, we estimate that the ratio of
the surface magnetic field strength on accreting BDs and stars is $\sim 0.6$. Since
T Tauri stars have in many cases magnetic fields of several kG \citep[e.g.][]{shk05}, this 
result indicates that young substellar objects may host magnetic fields in the range of 
1\,kG. Although the uncertainties in the given estimate are considerable (see above), we 
tentatively conclude that magnetic field strengths for BDs are probably in the same order 
of magnitude as for stars. 

\section{Conclusions}
\label{conc}

This paper contains the first spectroscopic variability study for young accreting brown 
dwarfs. We obtained, depending on the target, 4-13 high-resolution spectra using the 
Magellan/Clay 6.5\,m telescope, covering timescales from a few hours to several weeks. 
Our targets are six sub-stellar objects in nearby star-forming regions, for which signs 
of accretion had been noted in previous studies. We measured line widths for the 
most prominent emission lines, either related to accretion, winds, or both. In all spectra, 
the continuum is faint, but not seriously affected by incomplete background subtraction or 
accretion veiling.

All our targets are variable in the emission lines. The most dramatic changes are
seen for 2M1207, 2M1101, and ISO217, whereas the remaining targets ($\rho$Oph-32, 
ChaH$\alpha$-1, LS-RCrA\,1) show significantly less variability. The most prominent line 
in all cases is the H$\alpha$ feature. Our targets exhibit a variety of H$\alpha$ profiles,
similarly to accreting T Tauri stars. Two objects (2M1207, ISO217) have strongly asymmetric 
and double-peaked profiles at least in parts of the time series, produced by a superposition
of broad emission and either a red- (2M1207) or blue-shifted absorption feature. 2M1101 
also exhibits a weak red-shifted absorption feature in parts of our time series.

Strong changes in the H$\alpha$ 10\% width are seen in 2M1207 and 2M1101, indicating 
variations in the accretion rate on timescales of weeks (2M1207) and hours (2M1101). In 
both cases, this finding is supported by additional accretion indicators in the spectrum.
For 2M1101, the variations are consistent with a scenario of non-steady, clumpy accretion.
We also see evidence for strong variability in the outflow rate in at least three targets
(2M1101, $\rho$Oph-32, ISO217), particularly if we compare our results with literature data.
In two cases (2M1207, ISO217) there are indications for a coupling between infall and
outflow rate changes. In some cases the variability is best explained by rotational
modulation of the emission line flux. For LS-RCrA\,1, the variability characterictic
supports a close to edge-on view of the disk.

One important result from this study is a cautionary note: given that most accreting 
targets are strongly variable, studies of accretion-related properties
have to be based either on large samples or on time series. One example is the recently
found accretion rate vs. mass relationship, which shows a suspiciously large scatter. We 
demonstrate that it is possible to explain this noise by taking into account variability
information. 

For at least two out of six targets, we have to assume asymmetric accretion flows to interpret
the emission line shape and variations. This indirectly supports funneling of the accretion,
as predicted in the magnetospheric accretion scenario for T Tauri stars, and thus provides 
evidence that this scenario also applies to brown dwarfs. This implies the existence of 
large-scale magnetic fields, which may have $\sim$kG field strengths, in accreting substellar 
objects.

\acknowledgments
Some of the observations were carried out by Alexis Brandeker, whose help
is gratefully acknowledged. We thank the Las Campanas Observatory staff for 
their assistance. The constructive comments from an anonymous referee 
helped to improve the paper. This research was supported by an NSERC grant 
and University of Toronto startup funds to R.J.

\clearpage

\begin{deluxetable}{llccccc}
\tabletypesize{\scriptsize}
\tablecaption{Target properties \label{targets}}
\tablewidth{0pt}
\tablehead{\colhead{Full name} & \colhead{Short name} & \colhead{SpT\tablenotemark{a}} &
\colhead{$v_\mathrm{rad}$($\mathrm{kms^{-1}}$)\tablenotemark{a}}& 
\colhead{$v_\mathrm{rot}$($\mathrm{kms^{-1}}$)\tablenotemark{a}} &
\colhead{No.\tablenotemark{b}} &
\colhead{H$\alpha$VI\tablenotemark{c}}}
\startdata
2MASS J1207334-393254   & 2M1207         & M8     & 11.2       & 13  & 13  & 29 \\
2MASS J11013205-7718249 & 2M1101         & M8     & $\sim 15$\tablenotemark{d}  & -   & 7   & 49 \\
$\rho$Oph-ISO032	& $\rho$Oph-32   & M7.5   & $\sim 0$\tablenotemark{e} & -   & 5   & 17 \\
ChaH$\alpha$-1          & ChaH$\alpha$-1 & M7.5   & 15.5       & 7.6 & 5   & 19 \\	     
ChaI-ISO217		& ISO217         & M6.25  & $\sim 15$\tablenotemark{d}  & -   & 5   & 39 \\
LS-RCrA\,1              & LS-RCrA\,1     & M6.5   & $\sim 2$   & 18  & 4   & 16 \\
\enddata
\tablenotetext{a}{References for spectral properties: 
\citet{g02,mjb03,mlb05,ntc02,cnk00,jg01,j05,mjb05,l04,fc01,bmj04}}
\tablenotetext{b}{Number of spectra in the time series}
\tablenotetext{c}{H$\alpha$ variability index, as defined in Sect. \ref{Halpha}}
\tablenotetext{d}{Average radial velocity of ChaI brown dwarfs}
\tablenotetext{e}{$v_\mathrm{rad}$ derived in this paper}

\end{deluxetable}

\clearpage

\begin{deluxetable}{ccccccccc}
\tabletypesize{\scriptsize}
\tablecaption{Observing log and H$\alpha$ measurements \label{log}}
\tablewidth{0pt}
\tablehead{
\colhead{Date 2005} & \colhead{UT} & \colhead{Slit} & \colhead{Binning} & \colhead{Target} & \colhead{Exp.time (s)} & \colhead{H$\alpha$ EW(\AA)\tablenotemark{a}} & \colhead {10\% width($\mathrm{kms^{-1}}$)\tablenotemark{b}} & \colhead{Comment}}
\startdata
29/01 & 05:25 & 0\farcs35 & $1\times 1$   & 2M1207            & $1\times 900$  & 110 & 215 & \\
29/01 & 08:08 &           &	          & 2M1207 	      & $1\times 1200$ &  64 & 215 & \\
01/02 & 05:56 &           &	          & 2M1207 	      & $1\times 1200$ &  86 & 209 & \\

17/03 & 00:38 & 1\farcs0  & $2\times 2$   & 2M1207            & $3\times 600$  & 334 & 308 & double-peaked\\
      & 01:15 &           &               & 2M1101	      & $3\times 600$  &  22 & 131 & \\
      & 06:45 & 0\farcs7  &               & $\rho$Oph-32      & $3\times 1200$ &  88 & 270 & \\
      & 08:14 & 1\farcs0  &               & 2M1207	      & $3\times 600$  & 387 & 274 & \\
      & 08:52 &           &               & 2M1101	      & $3\times 600$  &  17 & 135 & \\
      & 09:28 &           &               & LS-RCrA\,1        & $2\times 600$  &  80 & 270 & \\
      
18/03 & 00:23 & 1\farcs0  & $2\times 2$ & 2M1207	      & $3\times 600$  & 259 & 291 & double-peaked\\
      & 01:14 &           &             & 2M1101              & $3\times 900$  &  12 & 122 & \\
      & 02:06 &           &             & ISO217              & $3\times 600$  & 109 & 359 & \\
      & 02:41 &           &             & ChaH$\alpha$-1      & $3\times 600$  &  72 & 139 & \\
      & 05:16 &           &             & ChaH$\alpha$-1      & $3\times 600$  &  74 & 135 & \\
      & 05:50 &           &             & ISO217              & $3\times 900$  & 142 & 376 & \\
      & 06:46 &           &             & $\rho$Oph-32        & $3\times 1200$ &  82 & 224 & \\
      & 08:03 &           &             & 2M1207	      & $3\times 600$  & 229 & 253 & \\
      & 08:42 &           &             & 2M1101	      & $3\times 600$  &  92 & 232 & accretion burst\\
      & 09:19 &           &             & LS-RCrA\,1          & $3\times 600$  &  44 & 266 & \\
      
19/03 & 23:52 & 1\farcs0  & $2\times 2$ & ChaH$\alpha$-1      & $3\times 600$  & 180 & 165& \\
      & 00:28 &           &             & 2M1207	      & $3\times 600$  & 279 & 283 & double-peaked\\
      & 02:06 &           &             & 2M1101              & $3\times 900$  &  85 & 224 & accretion burst\\
      & 03:07 &           &             & ISO217              & $3\times 600$  & 226 & 367& \\    
      & 03:57 &           &             & ChaH$\alpha$-1      & $3\times 600$  & 142 & 143 & \\
      & 04:32 &           &             & 2M1207	      & $3\times 600$  & 308 & 257 & double-peaked\\
      & 05:07 &           &             & ISO217              & $3\times 600$  & 219 & 363& \\
      & 06:06 &           &             & ChaH$\alpha$-1      & $2\times 600$  & 132 & 152 & \\
      & 06:37 &           &             & $\rho$Oph-32        & $3\times 1200$ & 100 & 224& \\
      & 07:50 &           &             & 2M1207	      & $3\times 600$  & 258 & 270 & \\
      & 08:27 &           &             & 2M1101	      & $3\times 900$  & 126 & 194 & accretion burst\\
      & 09:29 &           &             & LS-RCrA\,1          & $3\times 600$  & 125 & 291 & \\
      
27/03 & 03:56 & 0\farcs35 & $1\times 1$ & 2M1207              & $1\times 1800$ & 192 & 295 & double-peaked\\
      & 06:29 &           & $2\times 2$ & $\rho$Oph-32        & $1\times 1200$ &  85 & 241 & \\
28/03 & 04:46 & 0\farcs35 & $1\times 1$ & 2M1207              & $1\times 1800$ & 183 & 279 & double-peaked\\
29/03 & 04:38 & 0\farcs35 & $1\times 1$ & 2M1207              & $1\times 1800$ & 261 & 285 & double-peaked\\
      & 05:14 &           & $2\times 2$ & ISO217              & $1\times 1800$ &  73 & 342 & \\
      & 05:47 &           &             & 2M1101              & $1\times 1800$ &  13 & 127 & \\
      & 07:54 &           &             & $\rho$Oph-32        & $1\times 1800$ &  50 & 211 & \\
30/03 & 05:12 & 0\farcs35 & $1\times 1$ & 2M1207              & $1\times 1500$ & 322 & 304 & double-peaked\\ 
      & 08:11 &           & $2\times 2$ & LS-RCrA\,1          & $1\times 1800$ &  71 & 304 & \\
\enddata
\tablenotetext{a}{The average measurement error in EW is $\sim 1\,$\AA.}
 \tablenotetext{b}{The measurement error in 10\% width is $\sim 5\,\mathrm{kms^{-1}}$}
\end{deluxetable}

\clearpage

\clearpage

\clearpage

\clearpage

\clearpage

\end{document}